\documentclass[journal abbreviation, manuscript]{copernicus}

\nolinenumbers

\usepackage{rotating, float, caption}

\begin{document}

\title{Regridding Uncertainty for Statistical Downscaling of Solar Radiation}


\Author[1,2]{Maggie D.}{Bailey}
\Author[1]{Douglas}{Nychka}
\Author[2]{Manajit}{Sengupta}
\Author[2]{Aron}{Habte}
\Author[2]{Yu}{Xie}\
\Author[1]{Soutir}{Bandyopadhyay}

\affil[1]{Colorado School of Mines, 1500 Illinois St. Golden, Colorado 80401}
\affil[2]{National Renewable Energy Lab 15013 Denver West Parkway Golden, CO 80401}




\correspondence{Soutir Bandyopadhyay (sbandyopadhyay@mines.edu)}

\runningtitle{Regridding Uncertainty}

\runningauthor{Bailey et al.}

\received{}
\pubdiscuss{} 
\revised{}
\accepted{}
\published{}


\firstpage{1}

\maketitle

\begin{abstract}
Initial steps in statistical downscaling involve being able to compare observed data from regional climate models (RCMs). This prediction requires (1) regridding RCM output from their native grids and at differing spatial resolutions to a common grid in order to be comparable to observed data and (2) bias correcting RCM data, via quantile mapping, for example,  for future modeling and analysis. The uncertainty associated with (1) is not always considered for downstream operations in (2). This work examines this uncertainty, which is not often made available to the user of a regridded data product. This analysis is applied to RCM solar radiation data from the NA-CORDEX data archive and observed data from the National Solar Radiation Database housed at the National Renewable Energy Lab. A case study of the mentioned methods over California is presented.
\end{abstract}

\copyrightstatement{TEXT} 

\section{Introduction}

Earth system models and regional climate models  (RCM) are standard tools used to quantify and understand future changes in climate. These models represent geophysical variables on fixed grids and so a comparison among models, with observations, or with other data products must reconcile the differences between variables registered on one set of grid locations to another set.  {\it Regridding} is a ubiquitous preprocessing step for climate model analysis to interpolate from one gridded field to another.  Common grid interpolation methods include kriging, cokriging, bilinear interpolation, inverse distance weighting, and thin plate splines (see \cite{mcginnis2010interpolation} for more details).  The uncertainty in these statistical and numerical interpolations has been well-documented (\cite{phillips1996spatial, loghmari2018performance}). However, to our knowledge this uncertainty is rarely factored into the analysis when a regridded field is considered. In the worst case regridded fields are distributed without the metadatas acknowledging the transformation from their native grid. 
Moreover, when regridded variables are used for a subsequent analysis, biases can be introduced into statistical estimates. 

This work is motivated by the practical issue of inferring the distribution of solar radiation across space and over the seasonal cycle from simulations provided by RCMs. The overall goal is to create a solar radiation data product at a high spatial and temporal resolution that is suitable for siting new solar power generation facilities, such as photo-voltaic plants. Since these facilities may have a lifetime of 30 or more years, it is important to factor in regional changes in climate in site planning.  The projections from a multi-model ensemble of RCM projections can suggest the  potential impacts of a changing climate on power generation.  However, we anticipate biases in the  regional model simulations as well as the need to combine several models in an optimal way. The National Solar Radiation Database (NSRDB) is a high resolution, gridded data product that can be used as a standard for calibration under current climate and is a benchmark training and testing sample. The initial step then is to build a statistical model that relates the regional climate model data, forced by reanalysis, to  a "gold standard" solar radiation data product
(NSRDB). 

\begin{figure}[t]
    \centering
    \includegraphics[width=.4\textwidth]{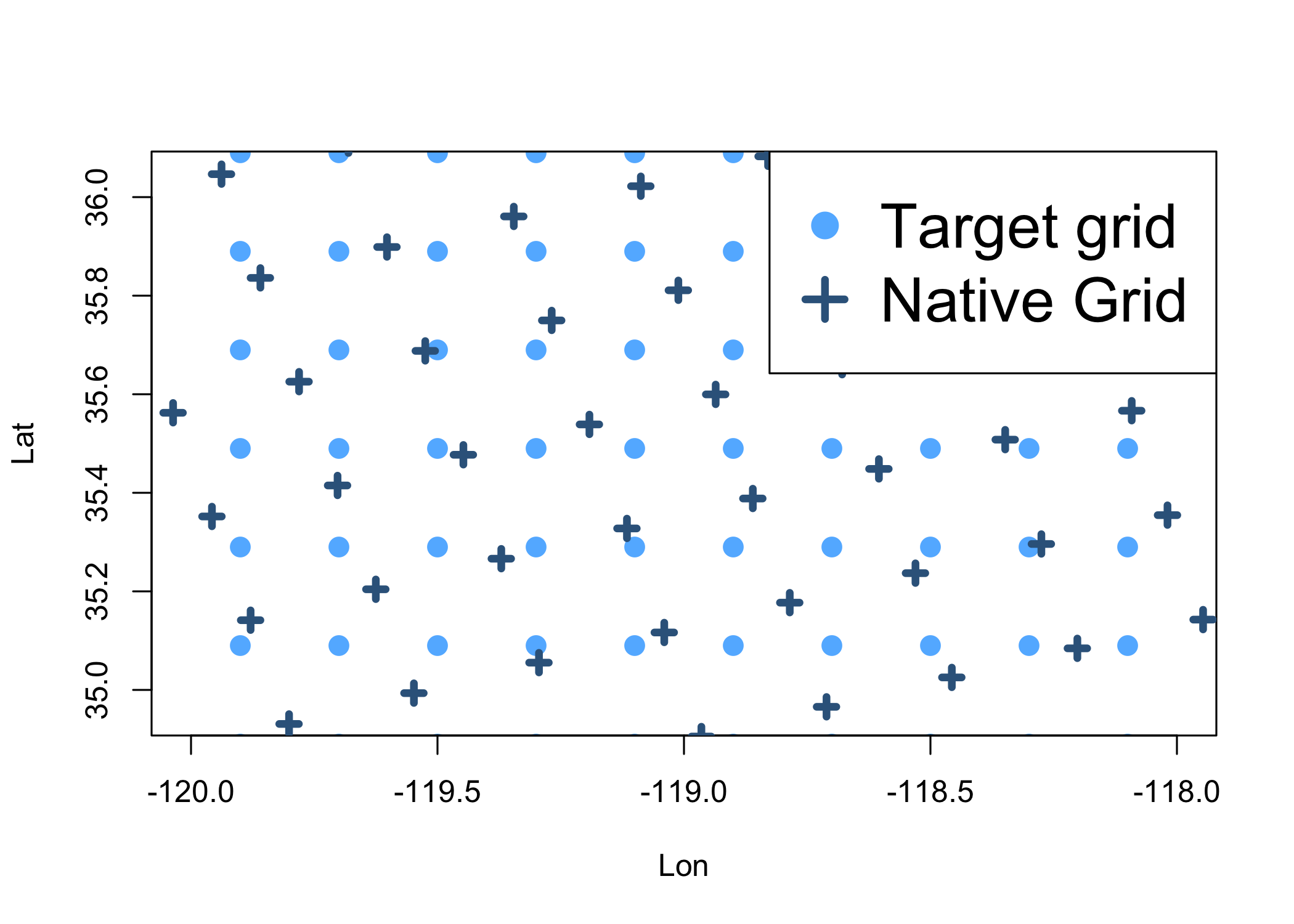}
    \includegraphics[width=.4\textwidth]{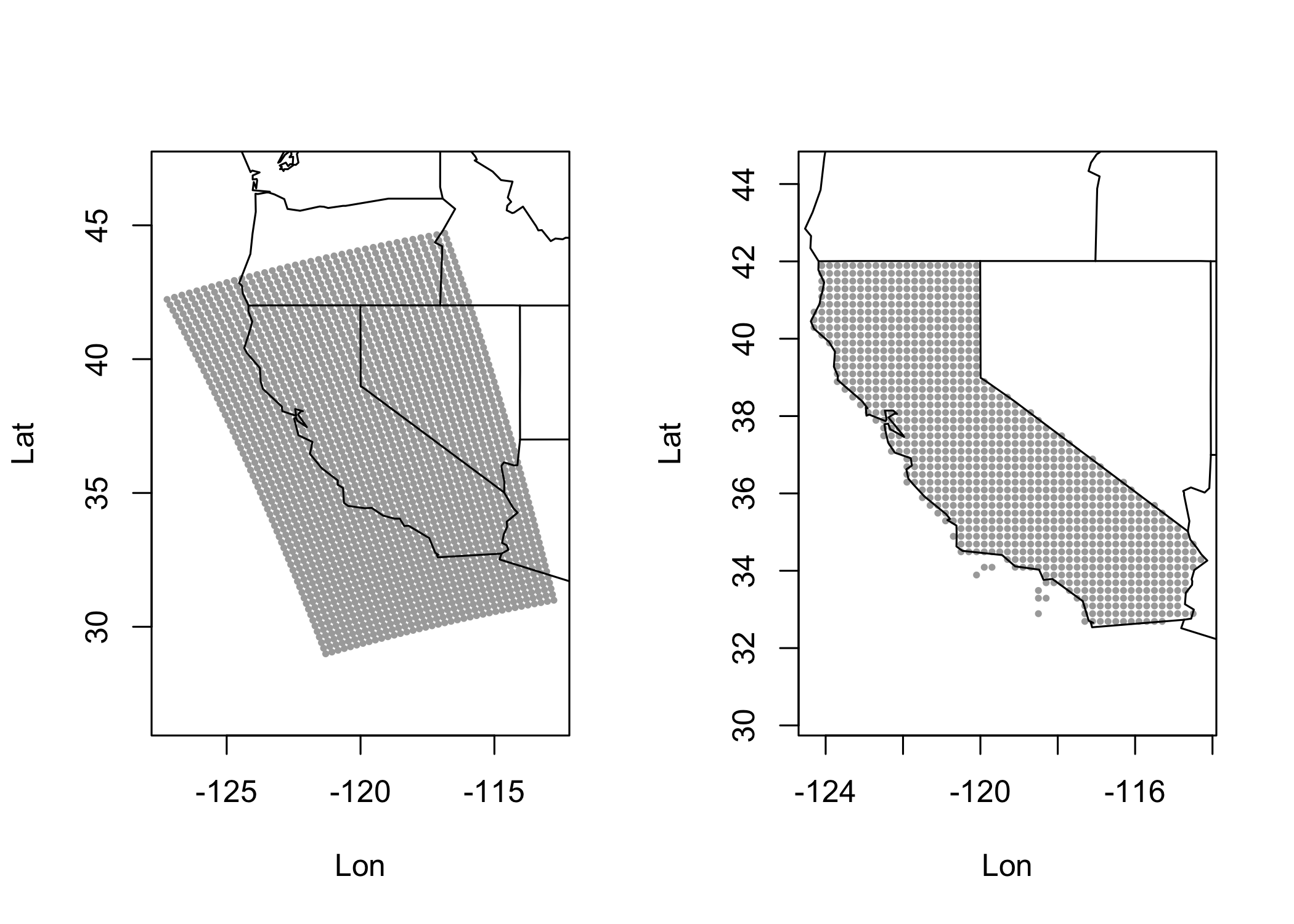}
    \caption{A comparison of differing grid projections. The left shows a close up of two differing grids on a 20km scale and a zoomed out projection on the right.}
    \label{fig:grid_comparison}
\end{figure}

Our  focus is on a linear model with NSRDB daily averages as the dependent variable and three RCMs as the independent variables for prediction.  The challenge is that the native grid for these models is not the same as the NSRDB, leading to the need for regridding. For illustration, consider Figure~\ref{fig:grid_comparison} showing grid projections from two different models. The differing projections result in an irregular pattern where some target grid locations are close to a native grid point while others fall in between grid locations. It is reasonable to assume that target grid locations that are close to native ones are more accurately interpolated than those further away and this varying uncertainty should be considered in the regridded version.

We contrast the approach of just using the regridded RCM fields as the regression predictors with an empirical  
Bayesian model that explicitly incorporates the mismatch between the RCM grids and the NSRDB grid. The  Bayesian approach takes advantage of recent tools in spatial statistics for conditional simulation of Gaussian processes and combines this  with classical Bayesian formulas for linear regression (\cite{Cressie2011StatisticsFS}). This strategy provides a 
simple framework to avoid the biases in an analysis based on a regridded estimate. Moreover this method can be 
extended to more sophicated prediction beyond a linear relationship. Our empirical Bayesian method uses the 
same spatial prediction model that would be used  in standard regridding but adds a step to generate 
conditional samples of the spatial fields. These realizations then  become the conditioned covariate in a Bayesian linear model and with a closed form expression for sampling the posterior of the regression parameters. The 
Bayesian approach is useful for determining unbiased estimates of the regression parameters. However, if the goal is simply prediction based on the linear model, we also show the standard regridding regression is appropriate for prediction, especially when prediction uncertainty is calibrated with a holdout sample. Therefore how this problem is tackled depends partly on the end goals of the analysis. 

We illustrate these ideas with an application to solar radiation prediction and these results are important in their own right. The analysis suggests the  limits of predictability of solar radiation based on  regional climate model simulations and also points to how the models may be biased relative to the NSRDB data set.

The uncertainty in the regridding process for solar radiation has not been given much attention as it relates to climate simulations, but there are many studies of  this  issue for  precipitation or temperature (\cite{chandler2022regridding, rajulapati2021perils}). \cite{mcginnis2010interpolation}
considered regridding error for RCMs in four regridding methods - nearest neighbors, bilinear interpolation, inverse distance weights, and thin plate splines for temperature and precipitation RCM data. The study found that thin plate splines performed the best of the four considered in terms of regridding but that the chosen interpolation method has a larger effect when considering local results as opposed to large-scale phenomenon across multiple models. Additionally, spurious extrapolation results need to be considered, particularly when considering extreme events, which temperature and precipitation analyses often do. The need for regridding was bypassed in \cite{harris2022multi} which proposes Neural-Network Gaussian Process Regression (NNGPR) for predicting temperature and precipitation reanalysis fields from ECMWF Reanalysis v5. The proposed method simultaneously downscales the same variables to RCM spatial levels using NA-CORDEX RCM data for validation. The method defines the downscaling pixel by pixel for the output grid by averaging the input climate model fields and defining a Gaussian process between the climate model fields and prediction point in the reanalysis field. Preliminary results from this study show marginal improvements over existing methods, including linear models, for combining climate models and poor uncertainty quantification skill, which is a direct focus of this study. Additionally, there are minimal metrics for uncertainty quantification of the method. While our method does not simultaneously downscale solar radiation data, this will be addressed in future work and the methods proposed in \cite{harris2022multi} could be used for a comparative analysis.

Effects of regridding on precipitation statistics has also been widely studied (\cite{accadia2003sensitivity, berndt2018spatial, ensor2008statistical, diaconescu2015remapping, rauscher2010resolution}). In particular, effects from regridding were found to have the largest impact at higher quantiles  (\cite{rajulapati2021perils}). The same study also found that the difference in precipitation statistics between the original and regridded data decreased with higher grid resolution, and vice-versa for lower resolution. This is also true at fine temporal scales (i.e. daily, sub-daily).

Our understanding is that there is a gap in the literature in analyzing the downstream effects on modelling after regridding, and for solar radiation data in particular. Note that this focus is related to the classic errors in variables models (\cite{whittemore1989errors}), where covariates are unknown or contain problematic data. Predictions based on the covariates containing error are reliable provided data with the errors is consistently used. However, inferring scientific relationships between the predictand and predictors is not reliable. Thus the focus in this study is whether final conclusions based on downstream modeling and evaluation of an RCM contribution to prediction skill will change when regridding uncertainty is taken into account.  As noted above, this study includes an analysis of possible effects using a Bayesian regression approach in order to quantify the uncertainty due to regridding.
In particular, we develop a Bayesian hierarchical model (BHM) as a complete description of the analysis uncertainties and then explain how simpler approaches result from approximations to this BHM (\cite{Cressie2011StatisticsFS}). Although we do not showcase a complete Bayesian analysis, we believe our approximate version is informative and is more easily implemented than the full Bayesian posterior computations. 


The rest of the article is organized as follows: Section~\ref{sec:data} introduces and describes the data utilized in this article, followed by an overview of the BHM in Section~\ref{sec:methods}; Section~\ref{sec:analysis_regrid_unc} details the method to analyze the uncertainty of regridding; Section~\ref{sec:results} shows results from this analysis; and finally, Section~\ref{sec:conclusion} concludes this work and discusses future directions.

\section{Data}
\label{sec:data}

\subsection{National Solar Radiation Database}

Modeled solar radiation data is part of the NSRDB, provided by the National Renewable Energy Lab (NREL) (see \cite{sengupta2018national} and the references therein for more details). The NSRDB is widely used by a variety of agencies, including local and federal governments, utility companies, and universities. Although primarily used for energy related applications, its uses have been extended to others such as the connection between solar exposure and cancer. The NSRDB contains hourly and half-hourly measurements for the three most common solar radiation variables: global horizontal irradiance (GHI), direct normal irradiance (DNI), and diffuse horizontal radiation (DHI), in units of $Wm^{-2}$. Solar radiation data is calculated using the Physical Solar Model which takes as input satellite measurements, cloud properties, and GHI derived from DNI and DHI. The data product for GHI and DNI has been validated and shown to be within 5\% and 10\%, respectively, when compared to surface observations. The NSRDB covers all of North America and surrounding countries for years 1998-2021 at the 4km resolution.  For the purposes of this study, however, the data set is averaged to a grid resolution of 20km to be comparable with climate model output at a similar resolution. The reader may note that the step of aggregation is related to the change of support problem (\cite{cressie1996change, gelfand2001change}). In aggregating from the 4km to the 20km grid for the NSRDB data set, we are hoping to avoid the change of support issue and to target the problem of regridding rather than the problem of downscaling. By using data that is on the 20km grid for the NSRDB, which closely aligns to the size of the RCM grid, we are isolating the problem of regridding and target the uncertainty related to this step only.

\subsection{Climate Model Database}

Climate model data is sourced from the North American CORDEX (NA-CORDEX) data archive, containing RCM output forced by various global climate models (GCMs) (\cite{mcginnis2021building}). The NA-CORDEX data archive contains many common climate variables, including surface downwelling shortwave radiation, at daily scales and sub-daily for some variables. Note that surface downwelling shortwave radiation is recorded as \textit{rsds} in NA-CORDEX but is equivalent to GHI and measured in the same units. Therefore, it will be referred to as ``GHI'' throughout this article. The archive includes ERA-Interim driven runs generally covering 1979-2014. GCM driven runs cover both historical periods (1949-2005) and future years (2006-2100) for various climate path scenarios and a model domain covering  all of the conterminous United States as well as most of Alaska, Canada, Mexico, and the Caribbean. 

The data chosen for this study will later be used in statistical downscaling operations for solar radiation for future years and RCMs for this study are chosen with this long-term goal in mind. The output chosen for this study are three ERA-Interim driven RCMs: Weather Research \& Forecasting Model (WRF), Canadian Regional Climate Model 4 (CanRCM4), and the fifth generation Canadian Regional Climate Model from the University of Quebec at Montreal (CRCM5-UQAM). These RCMs have desirable representative concentration pathways (RCPs) and grid resolution for future GCM driven runs. 

\section{Bayesian Hierarchical Model (BHM)}
\label{sec:methods}

A key part of this study is understanding the predictability of solar radiation based on RCM simulations. Here we organize the statistical assumptions as a BHM for clarity. This helps in tracing the Bayesian approximation used in our application and   the  standard approach based on regridding also follows as an additional approximation. 

Let $\{s_i\}$ be  the  NSRDB grid and  let  $y(s_i, t)$ denote average daily solar radiation reported at grid point $s_i$ and day $t$. For convenience, consider a single grid location  and let $\boldsymbol{y}^i= y(s_i, .) $ denote  the vector of all daily values at location $s_i$. In a similar way let $x_k(s_i,t)$ be the covariate value from the $k^{th}$ RCM at location $s_i$ and day $t$. The key assumption here is that this covariate makes sense at locations that are not part of the native grid of the RCM and so we define the ``correct'' covariate vector $X_k^i = x_k(s_i, .)$. With this assumption 
our main focus is on the linear model  given below, applied to available days and for each target grid location. 

\begin{equation}
    {\boldsymbol{y}}^i = \boldsymbol{\beta}_0 + \sum_{k=1}^M X_k^i \boldsymbol{\beta}_k +  S \boldsymbol{\gamma} + \varepsilon^i = 
    X^i \boldsymbol{\beta} + S \boldsymbol{\gamma} + \varepsilon^i 
    \label{eq:lin_model}
\end{equation}

where $\boldsymbol{y}^i$ represents the observed data from the NSRDB at the $i^{th}$ location,  $X^i$ is an $(M+1)$ column matrix formed from the constant term and $M$ many RCM covariates,  and $S$ is  matrix of seasonal covariates.  We will also assume $\varepsilon^i$ to be mean zero multivariate normal and variance $\sigma^2$. The estimated coefficients are the relative adjustments in matching the RCM simulations to observations. Overall the goal is to estimate the coefficients $\boldsymbol{\beta}$ and $\boldsymbol{\gamma}$ and use these in a prediction model for unobserved $\boldsymbol{y}^i$. Of course this would be a straightforward problem in linear regression if $X^i$ was known. However, the complicating issue is that $X^i$ must be inferred from other locations of the RCM grid. Moreover, we also note the complication that  $\varepsilon^i$ may exhibit dependence over time, which makes it more difficult to derive valid measures of uncertainty. This will be discussed in Sect.~\ref{sec:bayes_ts_addition}. The values of $\varepsilon^i$ will not show dependence over space but may show correlation at differing locations, $i \neq i^\prime$. As we are running this model from climate model reanalysis data, the spatial correlation of the errors should naturally be accounted for when $i\neq i^\prime$. 

\subsection{Observation and Process levels}

We assume a  stationary Gaussian process model for transformed RCM fields that are independent over time. 
Let $\{ u_j \} $ denote the native grid for the RCM simulated fields,  distinct in  the number and position from the NSRDB grid. 
The raw RCM solar variable, $ x$,  is transformed to $x^*$  with the  log/linear function according to 
\begin{equation*}
    x^* =  \Gamma(x) =
    \begin{cases}
        \log(x) & \text{if } x\leq \nu \\
        \log(\nu) + \frac{1}{\nu} (x- \nu) & \text{if } x> \nu 
    \end{cases}
    \label{eq:loglinear}
\end{equation*}
for a fixed value of $\nu$. 
We found this  transformation makes a Gaussian process assumption on $x^*_k(.,t)$ more reasonable and constrains predictions transformed back into the original scale to be positive. For $x>\nu$, $\Gamma(x)$  reverts to  a linear transformation and so retains interpretability for the larger and more relevant solar values.  In this study, $\nu$ is set to be the $20th$ percentile of the RCM values. The value of $\nu$ was chosen from a range of possible percentiles, spanning $[10th, 50th]$, as it resulted in summary statistics for the regridded values that were closest to the same statistics calculated from RCM data on the native grid. In this way, the regridded values are as similar as possible using $\nu=20th$ percentile than if a different value had been chosen. 

Given  these transformed fields we assume for fixed $t$, 
$x^*_k(s,t)$ will be a Gaussian process with mean function $\mu_k(s) = \mathbb{E}[x_k(s)]$ and  stationary, exponential covariance 
function:


\begin{equation}
    COV(  x^*_k(s,\cdot), x^*_k(s^\prime,\cdot)) = \rho_k \exp ( -\| s- s^\prime \|/ \theta_k),
    \label{eq:cov_gp}
\end{equation}

and for $t \neq t^\prime$

\begin{equation*}
    COV(  x^*_k(s,t), x^*_k(s^\prime ,t^\prime)) = 0,
\end{equation*}

that is, we are assuming no temporal correlation in this process. However, we will relax this assumption in Sect.~\ref{sec:bayes_ts_addition}. Finally, in the following it is convenient to let $x_u^t$ denote the $k^{th}$ RCM values on its native grid and for day $t$. With this setup we have the BHM defined in Table~\ref{tab:BHM}. Here the latent processes are independent Gaussian Processes with covariance given in Eq.~\ref{eq:cov_gp}. The joint distribution for this problem  is thus
 \begin{equation*}
      [ {\color{magenta} \boldsymbol{y}^i } |  \boldsymbol{\beta}, \boldsymbol{\gamma}, X^i, {\color{magenta} S^i}, \sigma^2] \quad [ X_1^i |  {\color{magenta} \{ x^*_1(u_j ,t) \} } , \rho_1 , \theta_1 ] \ldots 
 [ X_M^i | {\color{magenta}   \{ x^*_M(u_j , t) \} }, \rho_M , \theta_M ] \quad  [ \boldsymbol{\beta}, \boldsymbol{\gamma},  \sigma^2, \{ \rho_k\} , \{ \theta_k\} ] 
 \end{equation*}
where we have highlighted in color the terms based on observed data that are fixed in the Bayesian computation. 
The Bayesian formalism is to identify this 
 expression via Bayes theorem as proportional to the posterior density for the unknown parameters. Conceptually one would integrate this expression over $X_M^i$ to give a marginal posterior density in the statistical parameters. An exact integration, however, does not have a closed form and instead the standard approach is to sample from the posterior using Monte Carlo methods. 

\begin{table}[h]
    \centering
    \caption{BHM for including regridding uncertainty into the coefficient estimates for multi-model analysis. }
    \label{tab:BHM}
    \begin{tabular}{p{0.2\linewidth}  >{\centering\arraybackslash}p{0.55\linewidth}  p{0.25\linewidth} }
        \multirow{2}{5.5em}{\textbf{Observations} } & $[ {\boldsymbol{y}}^i |  \boldsymbol{\beta}, \boldsymbol{\gamma}, X^i, S^i, \sigma^2] \sim MN( X^i \boldsymbol{\beta} + S^i \boldsymbol{\gamma}, \sigma^2 I) $ & \textit{NSRDB} \\
          &  $[x_1^i | \rho_1, \theta_1, \mu_1]$, \ldots, $[x_M^i | \rho_M, \theta_M, \mu_M]$ & \textit{sampled RCMs on native grid and for all days} \\
         \hspace{3mm} & & \\
        \textbf{Latent process} & $ [\Gamma^{-1}(x^*_k(u,t))|\rho_k, \theta_k] $  for $k =1, \ldots, M$ and $t = 1, \ldots, T$ &  \\
        \hspace{3mm} \\
        \textbf{Priors on statistical parameters} & $[ \boldsymbol{\beta}, \boldsymbol{\gamma},  \sigma^2, \{ \rho_k\} , \{ \theta_k\} ] $ & 
    \end{tabular}
\end{table}


\subsection{Approximate BHM}
\label{sec:approx_bhm}
 From a Bayesian perspective this posterior is a complete characterization of the uncertainty in all unknown quantities. Unfortunately  in this case, as in many 
 BHMs, there is not a closed form for the normalized posterior and so one approximates this 
 distribution. In our case a complete sampling of the posterior is complicated by the fact that the posterior for the Gaussian process covariance parameters is coupled to the linear model through the RCM covariates. Because the linear model only depends on the RCM through its value at the observation grid one can break the sampling into two obvious steps and so doing  arrive at the usual strategy used for regridding.

\begin{description}
\item[Step 1] [Bayesian regridding] Sample $X_k^i $ from the density proportional to
\begin{equation*}
    [ X_k^i  |  {\color{magenta} \{ x^*_k(u_j ,t) \} } , \rho_k , \theta_k ] [\rho_k , \theta_k]
\end{equation*}

\item[Step 2] [Bayesian linear model] Sample  $  \boldsymbol{\beta}, \boldsymbol{\gamma},  \sigma^2 $  from the density proportional to 
\begin{equation*}
    [{\color{magenta} \boldsymbol{y}^i } |  \boldsymbol{\beta}, \boldsymbol{\gamma}, X^i, S^i, \sigma^2] [ \boldsymbol{\beta}, \boldsymbol{\gamma},  \sigma^2]
\end{equation*}

 \end{description}
 
The first step has been studied  by  several authors (\cite{Handcock1993ABA, finley2013spbayes}) and  is implemented with publicly available software and approximations for larger numbers of locations.  This step can 
be approximated further by an empirical Bayes assumption where the maximum likelihood estimates (MLE)  for $ 
\rho_k$, $\theta_k $, and $\mu_k$ are substituted into the conditional distribution. 
Explicitly this approximation leads one to sample  $X_k^i$  from a multivariate normal distribution 
\begin{equation*}
   X_k^i \sim MN( \mu_k^i, \Sigma_k^i).
\end{equation*}

Here $\mu_k^i$ is the conditional mean for the RCM field on the target grid {\it given} the values on the native grid 
and  {\it given} the covariate parameters at the MLE.  $ \Sigma_k^i$ is interpreted as the conditional covariance 
matrix describing the variation for the target grid values conditional on the native grid ones. 
Sampling from this distribution for fixed covariance parameters is referred to as conditional simulation in 
geostatistics and several computational algorithms have been developed for large problems.  We also note that the 
conditional mean vector, $\mu_k^i$, in this setting is the well-known ``Kriging'' spatial prediction from geostatistics.
Finally, in this step if one skips any sampling and sets $X_k^i \equiv \mu_k^i $ then  this is the standard practice for regridding. 


Several regridding methods were considered for this study, including thin plate splines and bilinear interpolation. The chosen method, Kriging with an exponential covariance function, performed the best when considering mean-squared error on test data. Because this study is focused on the uncertainty in this step itself and, more importantly, the downstream effects of the regridding step, a single regridding method was chosen. The differences between regridding and interpolation methods themselves are considered in other studies (\cite{mcginnis2010interpolation}).

The second step, conditional on having $X^i$ in hand is a standard Bayesian linear model. In our case, although not necessary,  we adopt a noninformative, uniform prior on $\boldsymbol{\beta},  \boldsymbol{\gamma}$, and $\log(\sigma^2)$ written as
\begin{equation*}
    [\boldsymbol{\beta},  \boldsymbol{\gamma}, \sigma^2 |X^i] \propto \sigma^{-2}
    \label{eq:prior}
\end{equation*}

giving a posterior of 
\begin{equation}
    [\boldsymbol{\beta}, \boldsymbol{\gamma}, \sigma^2 | \boldsymbol{y}^i, X^i] = [\boldsymbol{\beta}, \boldsymbol{\gamma} | \sigma^2 ,  \boldsymbol{y}^i, X^i] [ \sigma^2 |  \boldsymbol{y}^i, X^i]. 
    \label{eq:joint_post_1}
\end{equation}

For the first term   
\begin{equation}
    [\boldsymbol{\beta}, \boldsymbol{\gamma} | \sigma^2 ,  \boldsymbol{y}^i, X^i] \sim  MN( ( \hat{\boldsymbol{\beta}}, \hat{ \boldsymbol{\gamma}} ), (\sigma^2 \Omega) ^{-1} ),
    \label{eq:marg_beta_gamma}
\end{equation}

where  $\hat{\boldsymbol{\beta}}$ and $\hat{ \boldsymbol{\gamma}}$ are the ordinary least squares point estimates for the parameters and 
\begin{equation*}
    \Omega(X^i, S^i)  =  \left( \begin{array} {cc}
 (X^i)^TX^i &  (X^i)^T S^i \\
 
(S^i)^T X^i& (S^i)^T S^i \\
\end{array}\right).
\end{equation*}

The marginal posterior distribution for $\sigma^2$ are taken from the inverse Chi-square distribution:
\begin{equation*}
    [\sigma^2|\boldsymbol{y}^i, X^i  ] \sim  \text{Inv}-\chi^2(n-k,s^2),
    \label{eq:sigma_posterior}
\end{equation*}

where $s^2$ is the unbiased variance estimate also from ordinary least squares. 
The resulting posterior distributions for $\boldsymbol{\beta}$ and $\sigma^2$ can then be compared to the linear model 
estimates for each location from Eq.~\ref{eq:lin_model}. As a final approximation the Bayesian computation 
can be made more efficient by omitting the posterior uncertainty in $\sigma^2$ and substituting the point estimate into the first 
conditional density. 

In summary, starting with a general formulation of predicting a field based on misregistered spatial covariates we have detailed the series of approximations to identify the regridding algorithm. Here the RCM covariables are found by geostatistical prediction and substituted for the unknown fields values. The problem with this approach is highlighted by the  matrix $\Omega^{-1}$ in Eq.~\ref{eq:marg_beta_gamma} that determines $\hat{\beta}$ and its uncertainty and is a nonlinear function of  $X^i$. Thus substitution of a conditional mean of $X_i$ into this expression will not be equivalent to the better approximation afforded by the sampling in Step 2 above. In particular, $\Omega(X^i,S^i)^{-1}$ is a nonlinear function of $X^i$. The conditional mean of for the matrix inverse is not the same as substituting in the conditional mean for $X^i$, and this difference suggests the BHM is the correct approach for quantifying the uncertainty calculated from $\Omega(\cdot)$.

\subsection{Autoreggression Component to BHM}\label{sec:bayes_ts_addition}

It's worth mentioning that, in the above analysis, there is no time series addition to explain possible autocorrelation in the residuals. However, we determined a temporal component may not be necessary through fitting several auto regressive moving average models of order $p$ and $q$ (i.e., ARMA$(p,q)$) and assessing resulting AIC and BIC values. Across the four months considered (February, May, August, November), ARMA(0,0) was largely the best model according to both AIC and BIC. However, in August a MA(2) model had the lowest AIC. Across all months, the model with the second lowest AIC or BIC was frequently a MA(2) or MA(1), followed by an AR(1) model. Although not crucial for our case study for completeness, we have included below how the analysis would change with the addition of a time series component for the Bayesian analysis. 

With a time series component for $\varepsilon^i$ in Eq.~\ref{eq:lin_model}, the final prediction results will be similar to those from the model in the previous section but the prediction variance will be affected by the inclusion of an ARMA$(p,q)$ model. Since single parameter models (i.e. AR($p$) or MA($q$)) had lower AIC and BIC more frequently than higher order models (i.e. ARMA($p,q$)), the time series parameter for both MA and AR models will be denoted generically by $\eta$ in the following. 

The joint posterior in Eq.~\ref{eq:joint_post_1} becomes 
\begin{equation*}
    [\boldsymbol{\beta}, \gamma, \sigma^2, \eta|\boldsymbol{y}^i, X^i] = [\boldsymbol{y}^i|\boldsymbol{\beta}, X^i, \sigma^2, \gamma, \eta][\boldsymbol{\beta}][\sigma^2][\eta].
\end{equation*}

Then, the joint density conditional on all parameters is
\begin{equation*}
        [\boldsymbol{y}^i|X^i, \boldsymbol{\beta}, \sigma^2, \eta] \sim MN((\hat{\boldsymbol{\beta}}, \hat{\gamma}),  \sigma^2 \Omega(\eta)) 
\end{equation*}

where now, the covariance matrix $\sigma^2\Omega(\eta)$ depends on the autocorrelation parameter(s) $\eta$. This is useful for single parameter time series models, such as AR(1) or MA(1), as well higher order models ARMA($p,q$) that are also causal and therefore can be written in an MA($\infty$) representation:
\begin{equation*}
    X_t = \sum_{j = 0}^\infty a_j w_{t,j} \cdots
\end{equation*}

and it can be shown that the MA($\infty$) process has the  autocovariance function
\begin{equation*}
    COV(X_t, X_{t+h}) = \gamma(h) = \sigma^2 \sum_{j=0}^\infty a_j a_{j+|h|}
\end{equation*}

with the parameters $a_j's$ depending on $p$ and $q$ from the AR and MA process, respectively.    

We now find the marginal posterior for $[\eta|\boldsymbol{y}^i, X^i]$. This can be done by averaging over $\boldsymbol{\beta}, \sigma^2$:
\begin{equation*}
        [\eta|\boldsymbol{y}^i, X^i] \propto \int [\boldsymbol{y}^i, X^i|\boldsymbol{\beta}, \sigma^2, \eta][\boldsymbol{\beta}, \sigma^2, \eta]d\boldsymbol{\beta} d\sigma^2.
\end{equation*}

After some algebra, $[\eta|\boldsymbol{y}^i, X^i]$ remarkably has a closed form found through this integration
\begin{equation*}                    [\eta|\boldsymbol{y}^i, X^i] \propto \frac{1}{(SS)^{(n-p)/2+1}}\frac{|X^T\Omega(\eta)^{-1}X|^{1/2}}{|\Omega(\eta)|^{1/2}}
\end{equation*}

where $SS = \sum_{i = 1}^n(y_i - \hat{f}(X^i))^2$ is the residual sums of squares found by fitting the linear model to $\boldsymbol{y}^i$ conditional on $X^i$. With the marginal posterior distribution for $\eta$ readily available, one samples values of $\eta$ and continues with the analysis as outlined previously. Section~\ref{sec:approx_bhm} can be done by transforming the data and response through multiplying by $\Omega(\eta)^{-1/2}$, converting to an ordinary least-squares problem.

\section{Solar Radiation Example}
\label{sec:analysis_regrid_unc}

The model described in Sect.~\ref{sec:methods} was fit once for each location in a subset area of California, shown in the far right bottom panel of Figure~\ref{fig:loc1_comparison}, which includes coastal and inland areas. Additionally, the model was fit for four separate months (February, May, August, and November) across all years of overlapping data (1998-2009). Initially, all covariates are included in the model. However, not all covariates were found to be significant. In particular, the CanRCM4.ERA-Int was found to hold no significance for the months of February, May, and November for most locations and no significance for about half of the locations in the month of August. The seasonal covariate held no significance for any of the months, which is expected since the data was subset into a single month from each season. Because of this, the seasonal covariate was removed for all four months.

\subsection{Posterior Distribution of Model Coefficients}

Often, the regridded data set resulting from kriging is used as the ground truth for further analysis. This section outlines the method for analyzing the uncertainty associated with the regridding and linear model prediction step and downstream effects by generating draws from the posterior distributions of each $\beta$ in Eq.~\ref{eq:lin_model}.

The study design is as follows. For each RCM, 100 independent draws are made from the conditional distribution onto the NSRDB grid. This is \textbf{Step 1} outlined in Sect.~\ref{sec:approx_bhm}.  For each of these simulations, 50 draws from the posterior distribution for $\beta$ and $\sigma^2$ are taken resulting in 5000 posterior draws. This is \textbf{Step 2} in Sect.~\ref{sec:approx_bhm}. Since each conditional simulation is equally likely, the posterior draws are aggregated and summarized together. This is done for each of the 114 locations in the study subset. The 95\% credible intervals for the posterior distribution and the 95\% confidence interval for each $\hat{\beta}$ are also considered. The 95\% credible intervals from the posterior distribution for the linear model parameters are determined by taking the 2.5 and 97.5 sample quantiles across the posterior samples.

\subsection{Prediction Coverage}

This study also considers the coverage of the posterior predictions, i.e. the fraction of days where the prediction interval contains the actual value observed. This portion of the analysis holds out a single year of data as a testing set, using the remaining years as a training set. There are 12 years of overlapping data resulting in 12 out-of-sample prediction results. The final coverage is the average coverage for the  12 folds.

\section{Results}
\label{sec:results}

The results presented here summarize the metrics outlined in Sect.~\ref{sec:analysis_regrid_unc}. As the true coefficients are not known, we have supplemented the analysis with a simulation study. The design and results for this study are described in given in Appendix~\ref{sec:appendix_a1}.

\subsection{Posterior Distribution of Model Coefficients}

\begin{figure}[t]
    \centering
    \includegraphics[width=\textwidth]{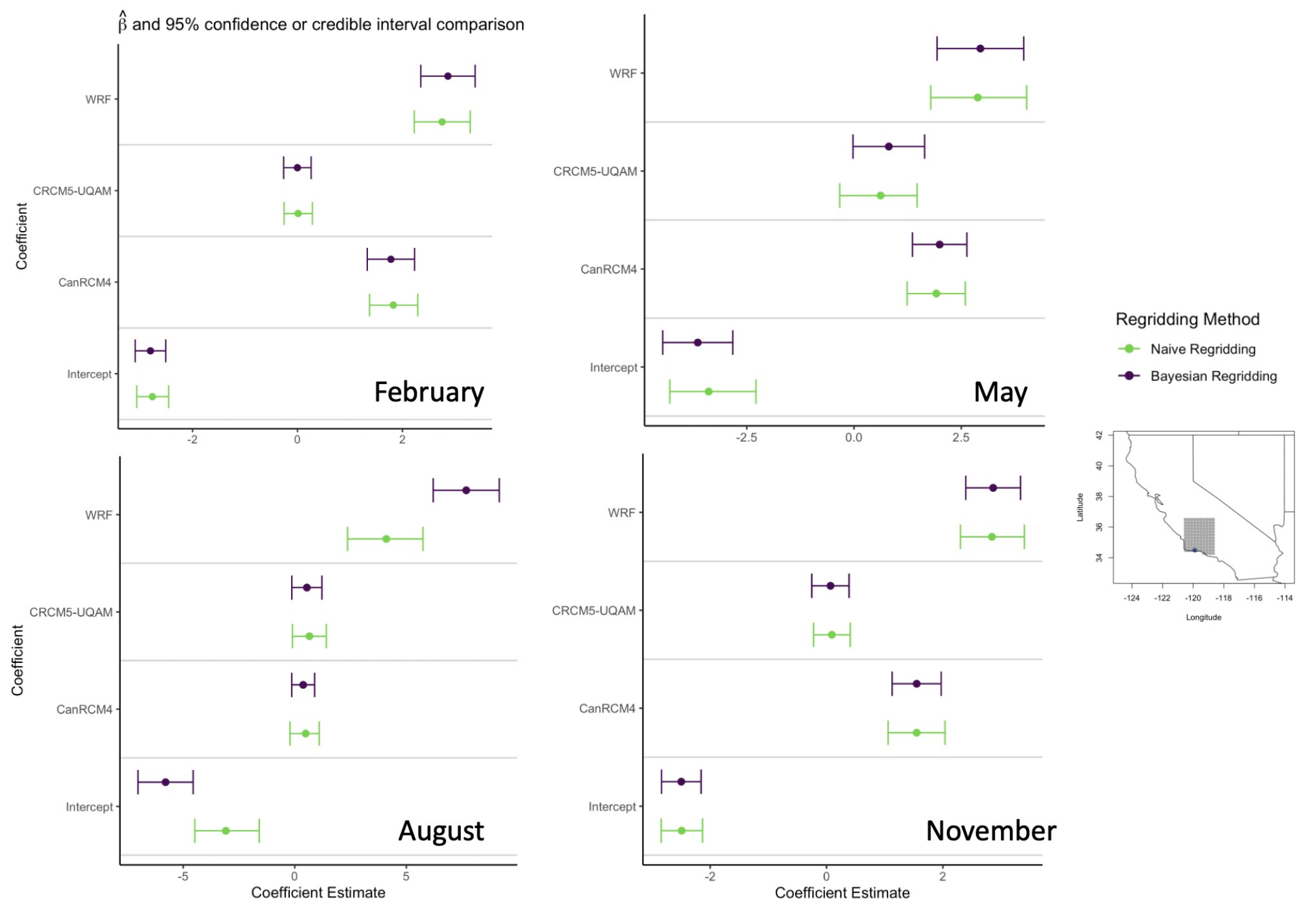}
    \caption{Posterior predictions for each coefficient compared to the naive regridding estimates for a particular location in California for February, May, August, and November (1998-2009). The solid dots represent the point estimate for the naive regridding method and the median value of posterior distribution from the Bayesian method. The whiskers represent the 95\% credible and confidence intervals for the posterior distribution and the naive regridding estimates, respectively.}
    \label{fig:loc1_comparison}
\end{figure}

Resulting parameter estimates from the posterior distribution vary by location and coefficient. Here we will refer to parameter bias as the difference between the naive estimate and that based on the Bayesian analysis. In general, the naive regridding model coefficient estimates are within the 95\% credible intervals of the posterior distributions for the respective coefficient. An example of the distributions compared to the naive regridding estimate can be seen in Figure~\ref{fig:loc1_comparison} for a location near the coastline of California across four different months. The green lines represent the naive regridding method and the purple the Bayesian regridding method. In general, there is strong agreement between the two methods in both the point or median coefficient estimate as well as the confidence or credible intervals suggesting that incorporating the uncertainty associated with the regridding step has little effect on model estimates. However, in the month of August (bottom left plot) we see a case for the WRF coefficient where the methods do not agree and this bias is off-set by the intercept estimate. This bias in the WRF coefficient was seen across many locations for the month of August. 

\begin{figure}[h]
    \centering
    \includegraphics[width=\textwidth]{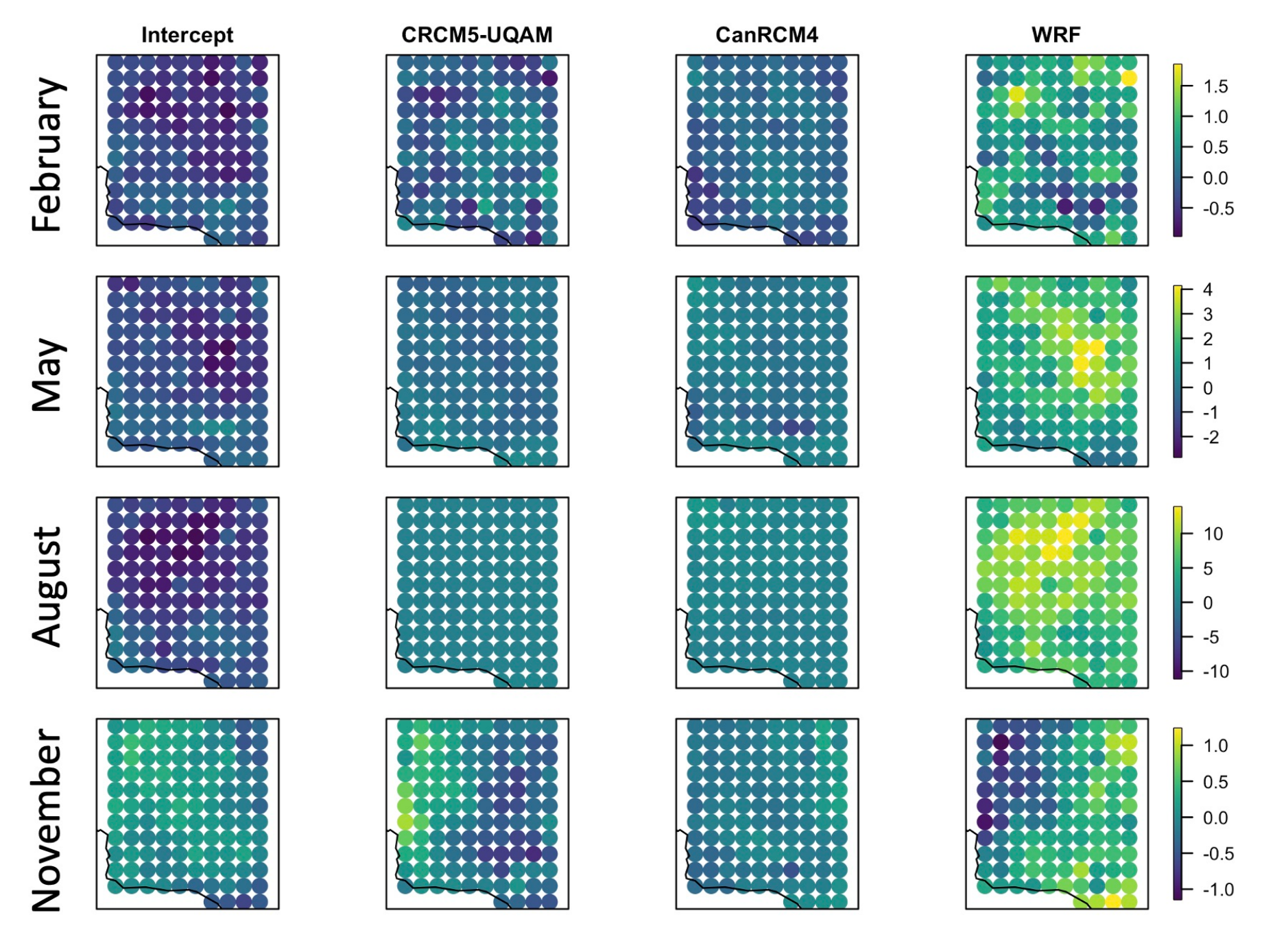}
    \caption{Average bias by location between the naive regridding estimate and the median of all posterior distributions for February, May, August, and November, top to bottom, respectively.}
    \label{fig:coeff_bias}
\end{figure}

For the entire area considered, the average bias by location is shown in Figure~\ref{fig:coeff_bias}. The bias is calculated as the BHM estimate subtracted from the naive regridding estimate. Values close to zero indicate little difference between the two methods. Negative values indicate that the BHM is giving a stronger weight to the model.  The spatial patterns of the bias are most pronounced for the month of November and also large for the month of August. In November, the average bias between the CRCM5-UQAM and WRF coefficient are spatially opposite in their signs but both hover around zero. Here, we can see that the naive method and BHM most disagree for the WRF coefficient in the month of August, with the naive method resulting in a much higher weight for WRF compared to the BHM. For additional reference, the estimated coefficient estimates and standard errors are provided in Appendix~\ref{sec:appendix_b1}. 


\subsection{Prediction Coverage and Error Comparison}

\begin{figure}
    \centering
    \includegraphics[width=.22\textwidth]{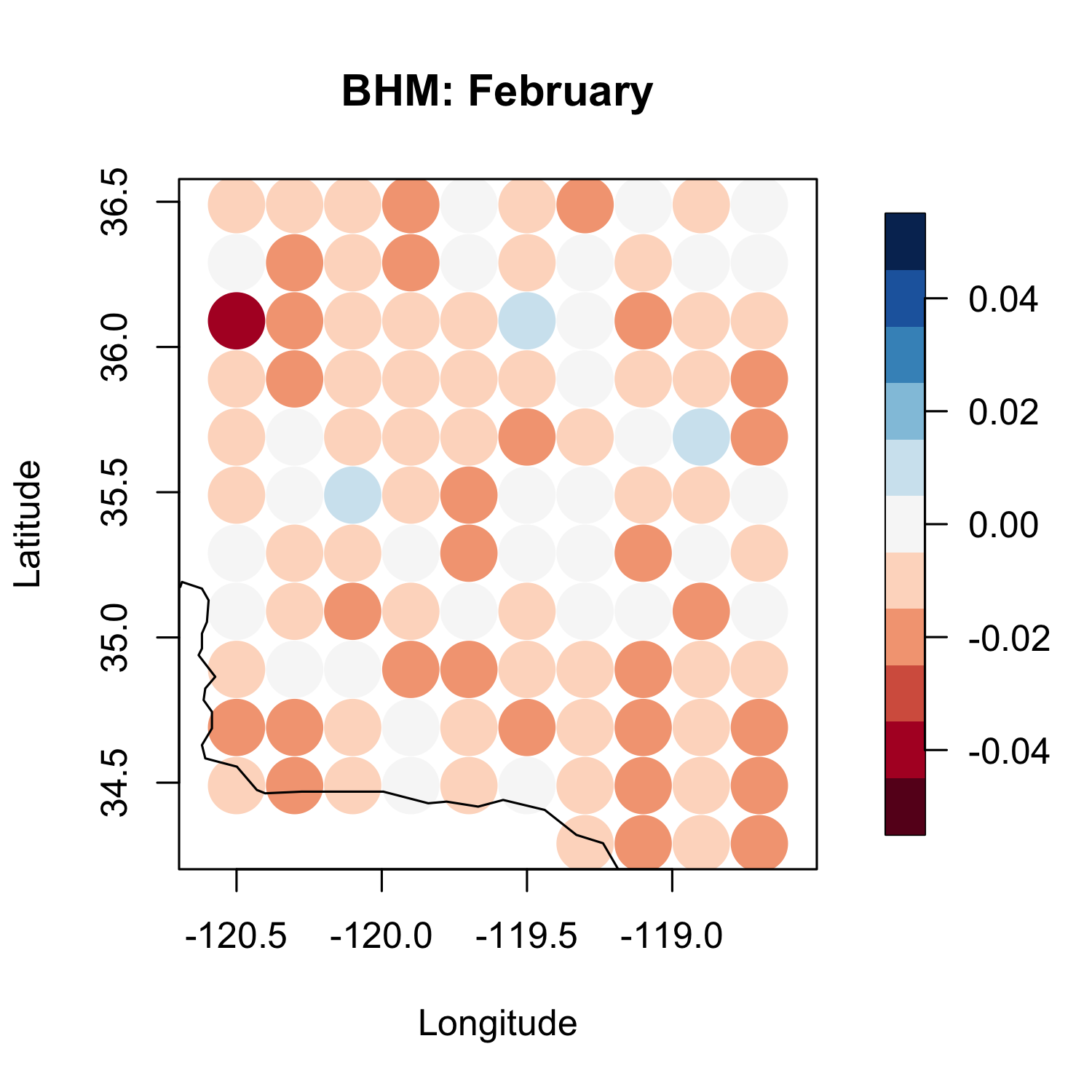} 
    \includegraphics[width=.22\textwidth]{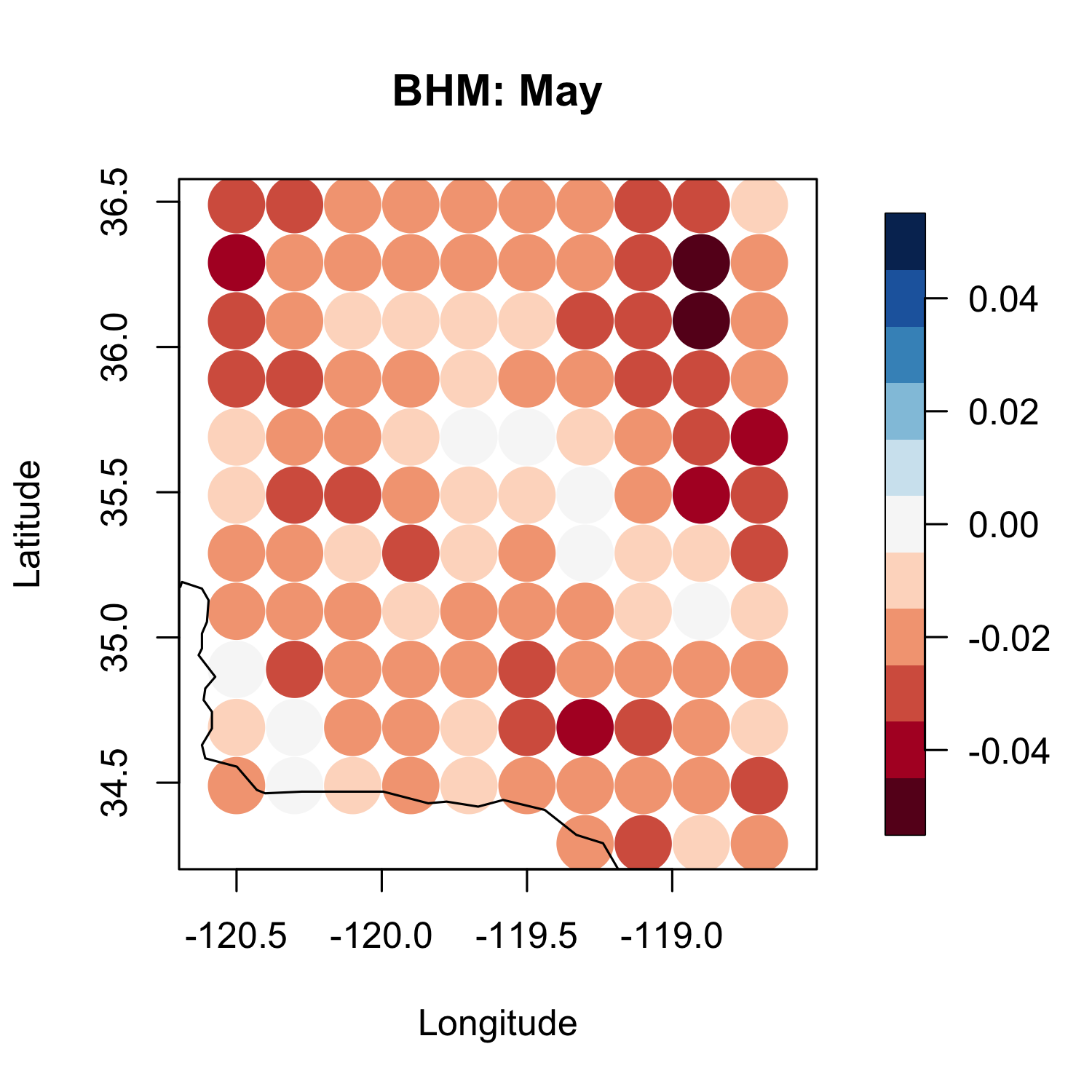} 
    \includegraphics[width=.22\textwidth]{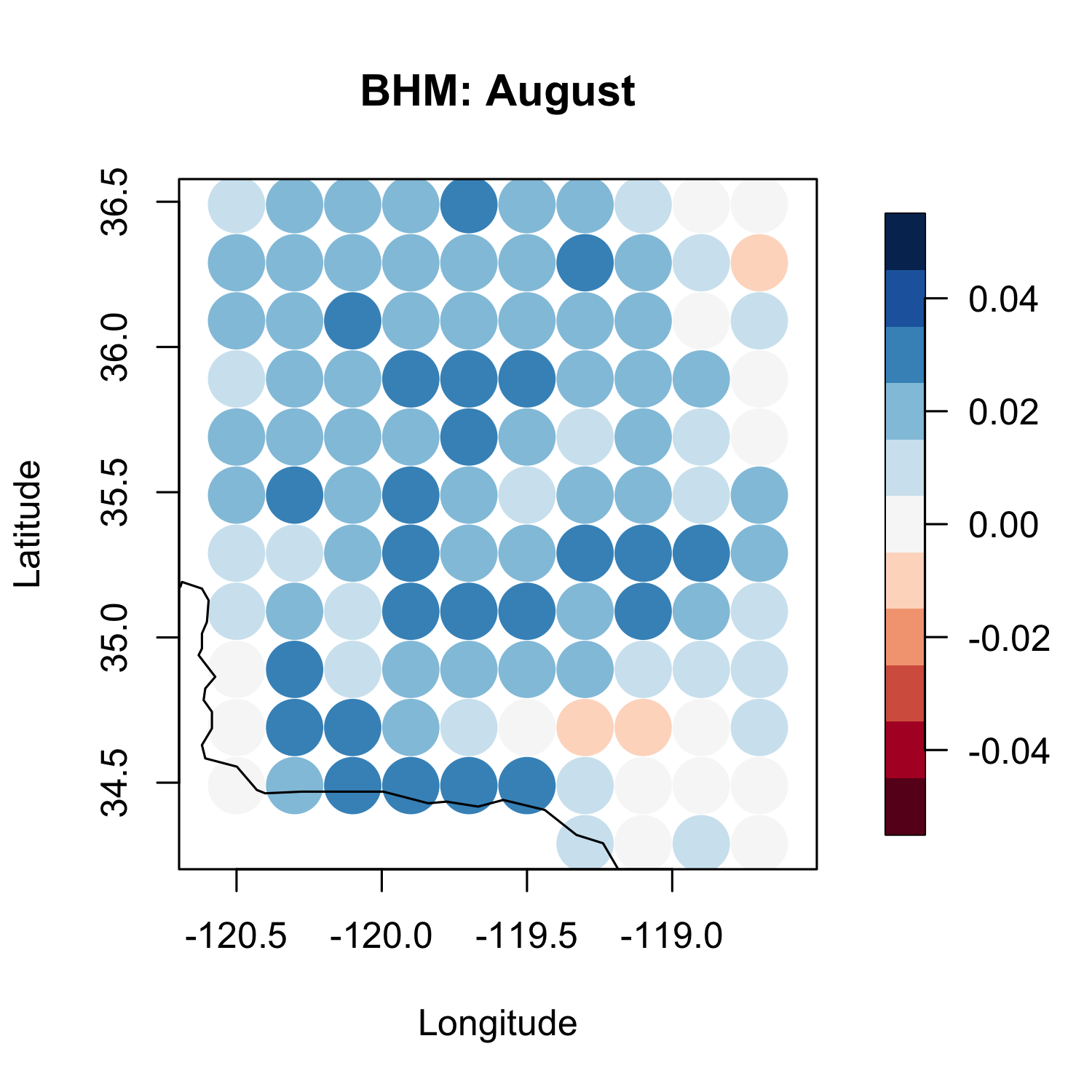} 
    \includegraphics[width=.22\textwidth]{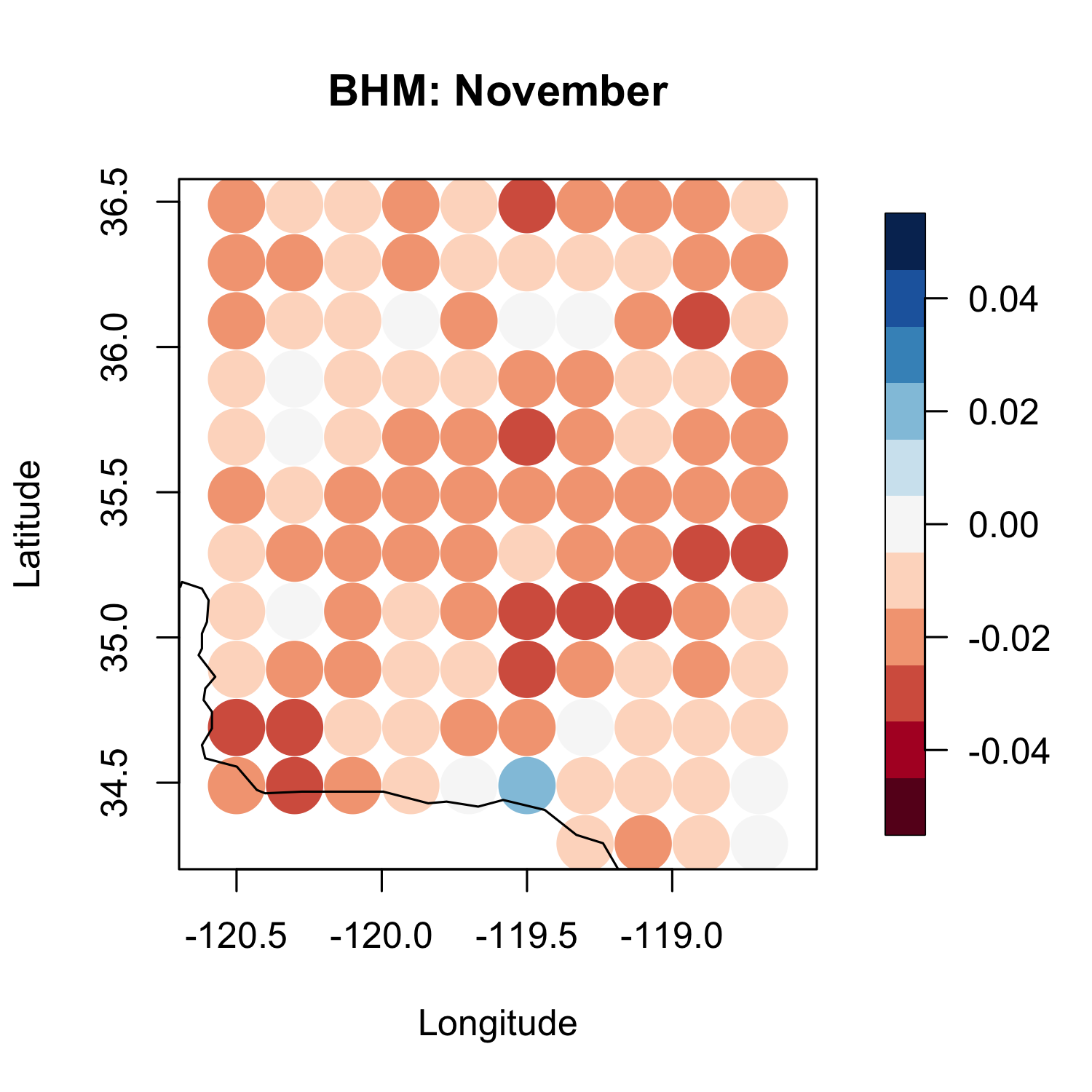}

    \includegraphics[width=.22\textwidth]{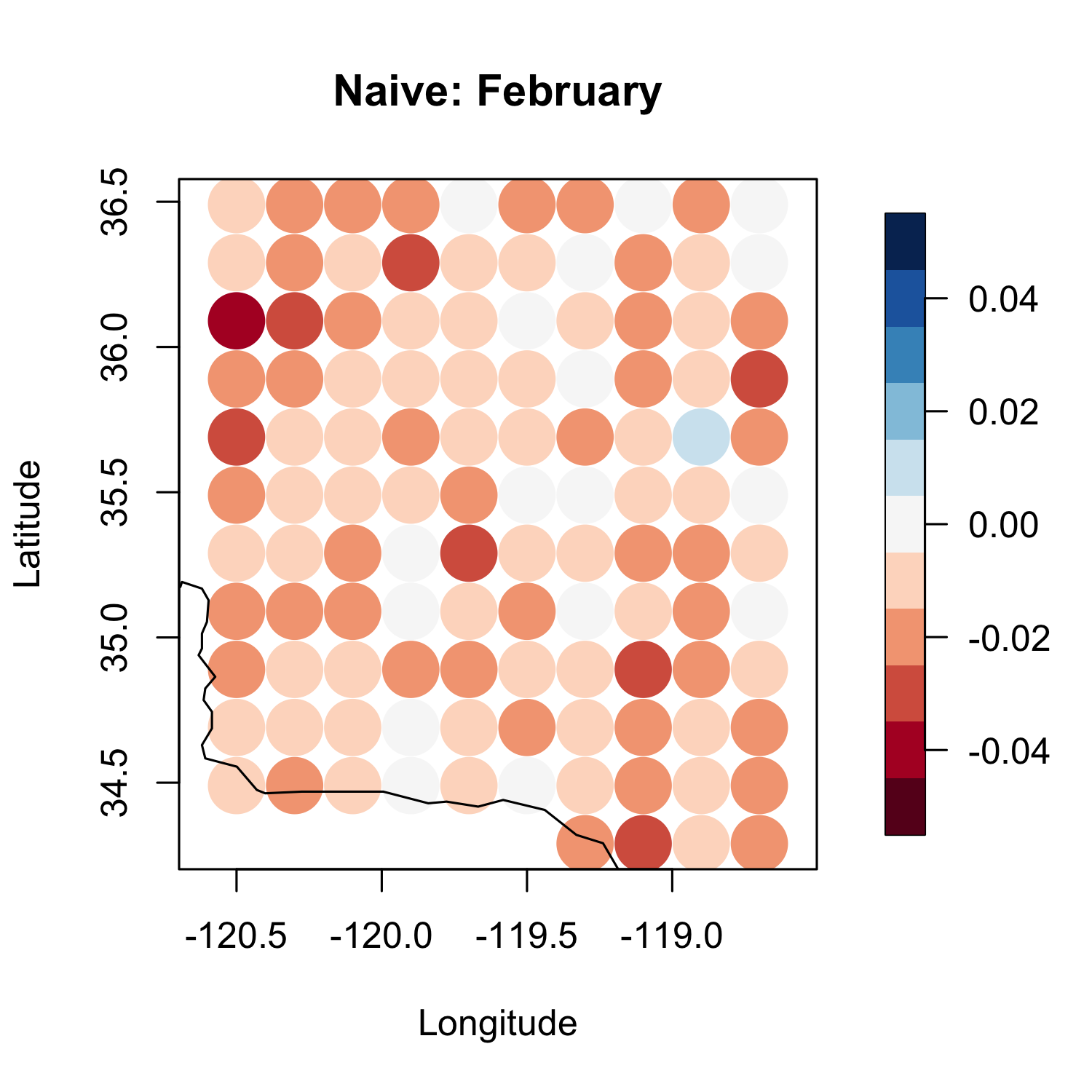}
    \includegraphics[width=.22\textwidth]{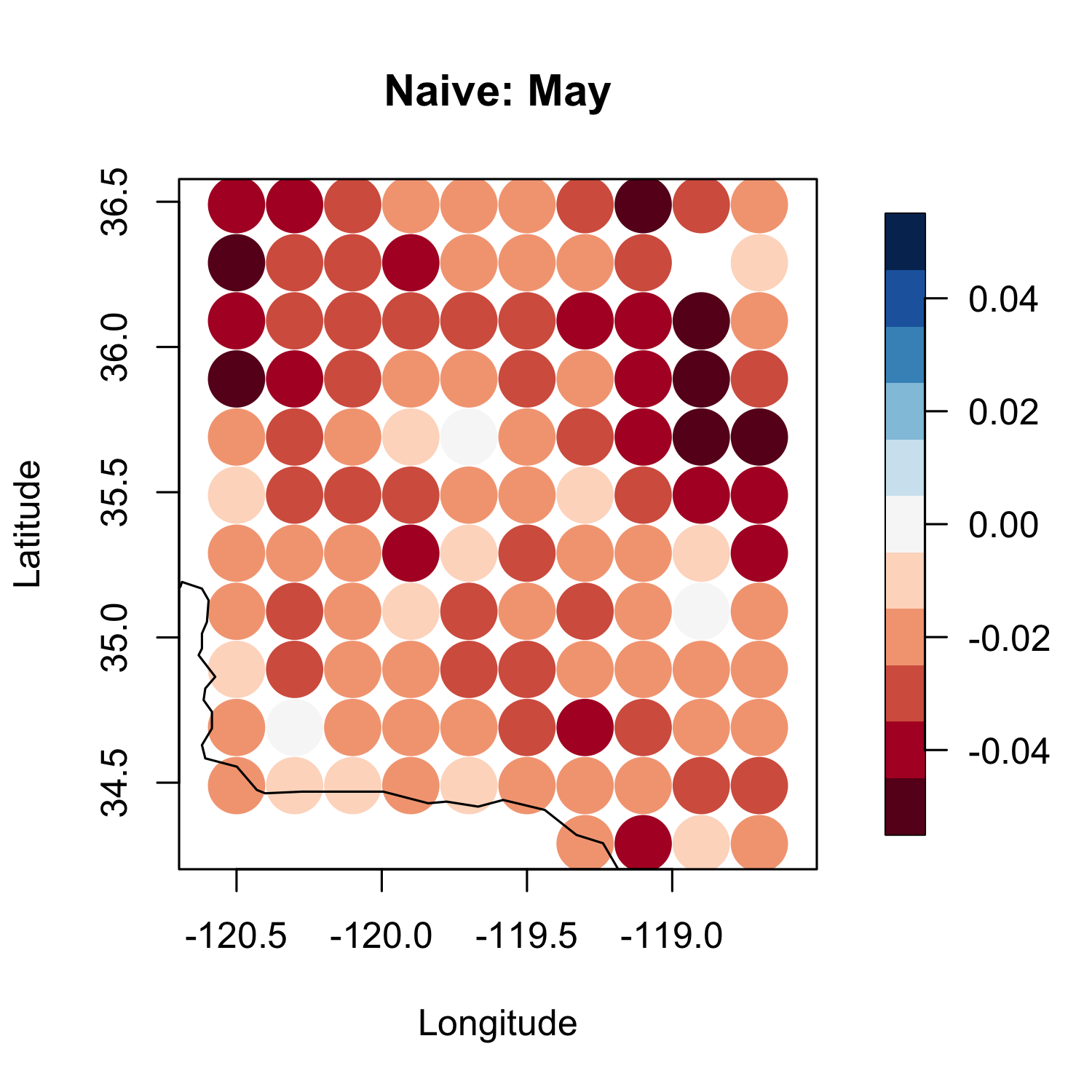}
    \includegraphics[width=.22\textwidth]{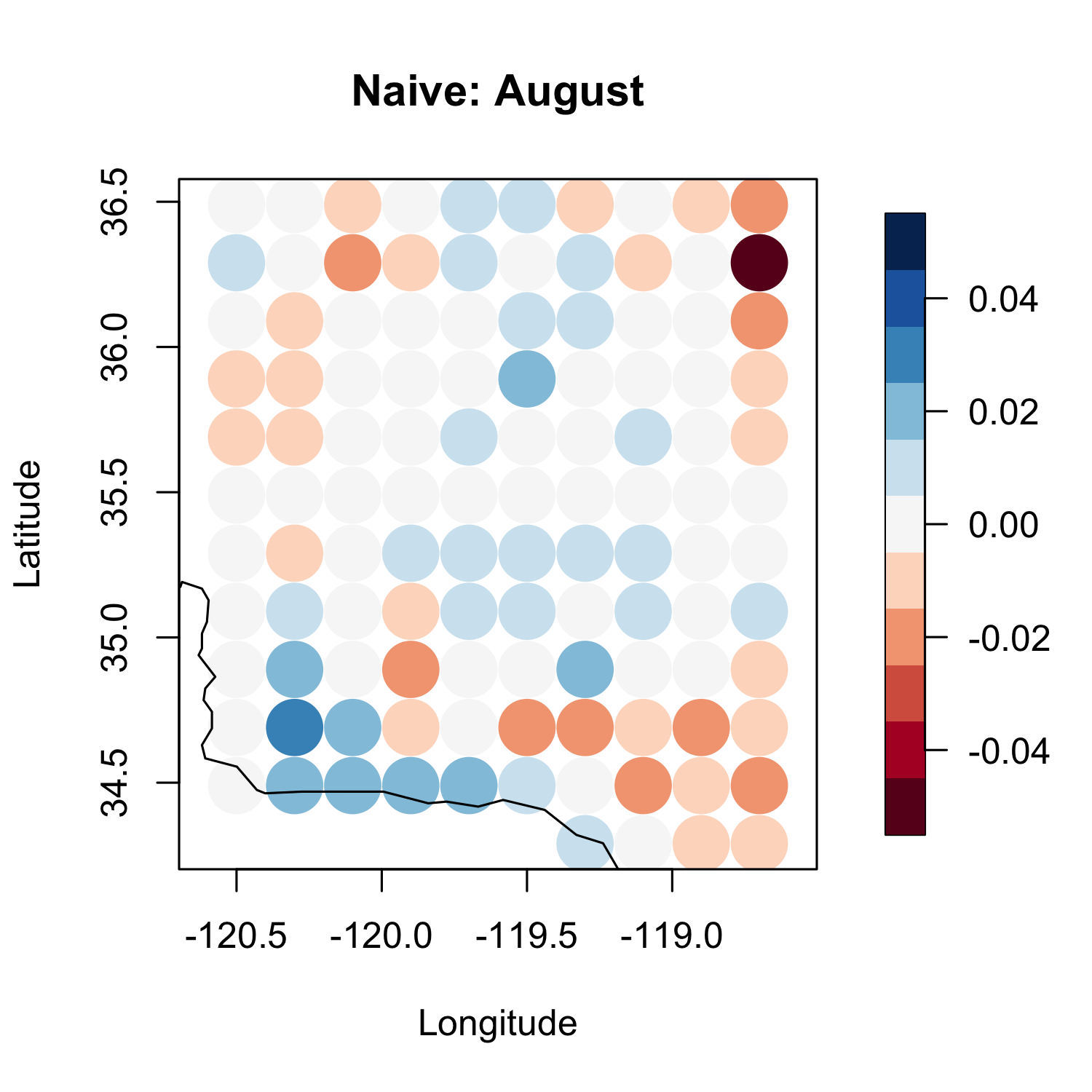}
    \includegraphics[width=.22\textwidth]{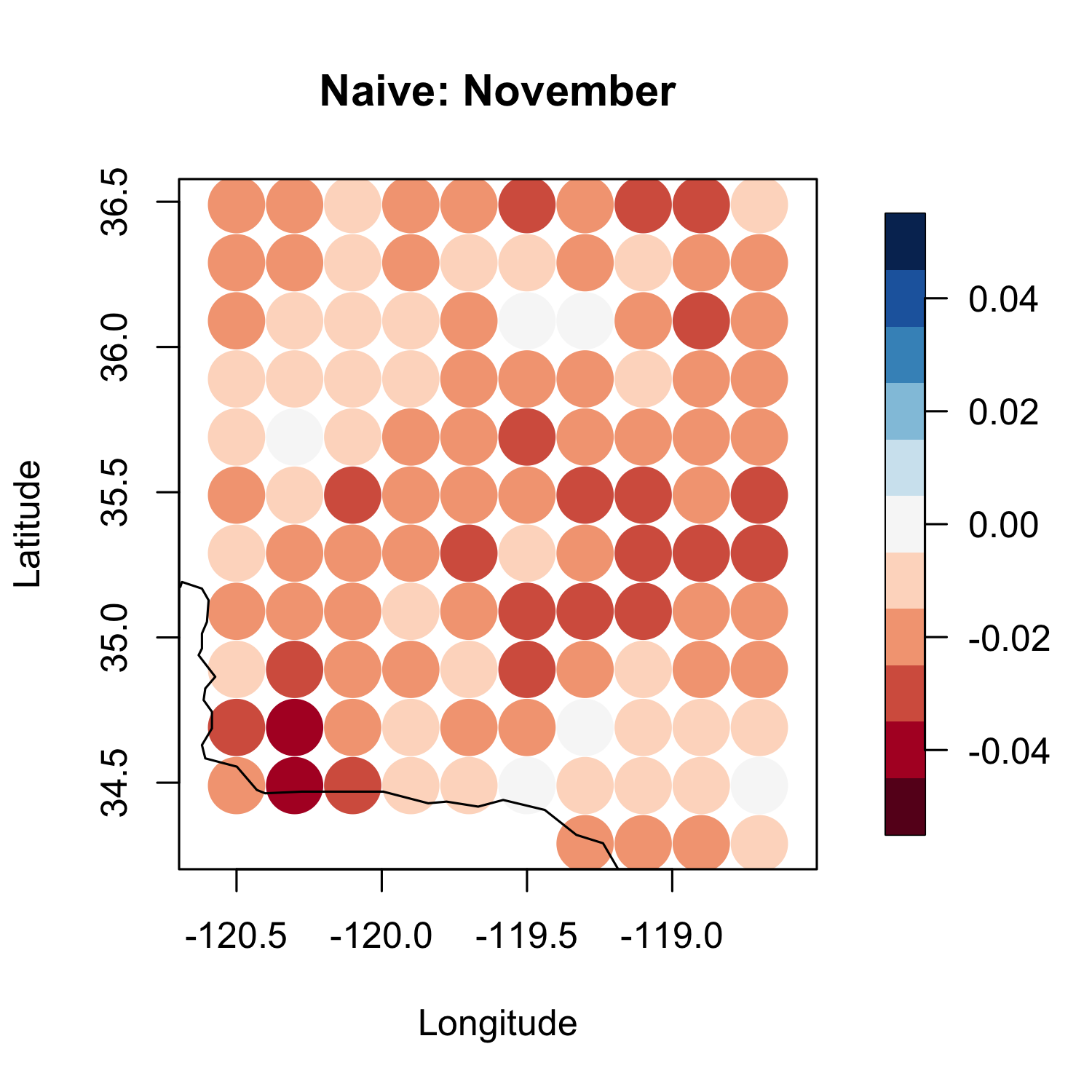}
    
    \caption{Coverage probability shown as the difference from the nominal level (0.95) by location after holding out each year from 1998-2009 individually for the four months considered. Bayesian model results are in the top row and naive regridding results in the bottom row. Coverage is similar between the two models, hovering around 0.95, and is slightly higher in August.}
    \label{fig:coverage_comparison}
\end{figure}

The prediction coverage of the naive regridding is calculated as the percentage of observations that are within the prediction intervals of the linear model. This is calculated by location for each of the four months considered. A similar method is implemented to calculate the coverage resulting from the BHM. We show results for the fourth months in Figure~\ref{fig:coverage_comparison}. Note that in the figure shown, the percent coverage reported is an average for holding out each year and shown as the difference from the nominal level of 0.95. We see similar results for the out of sample coverage compared to the naive regridding.

\begin{figure}[h]
    \centering
    \includegraphics[width=.22\textwidth]{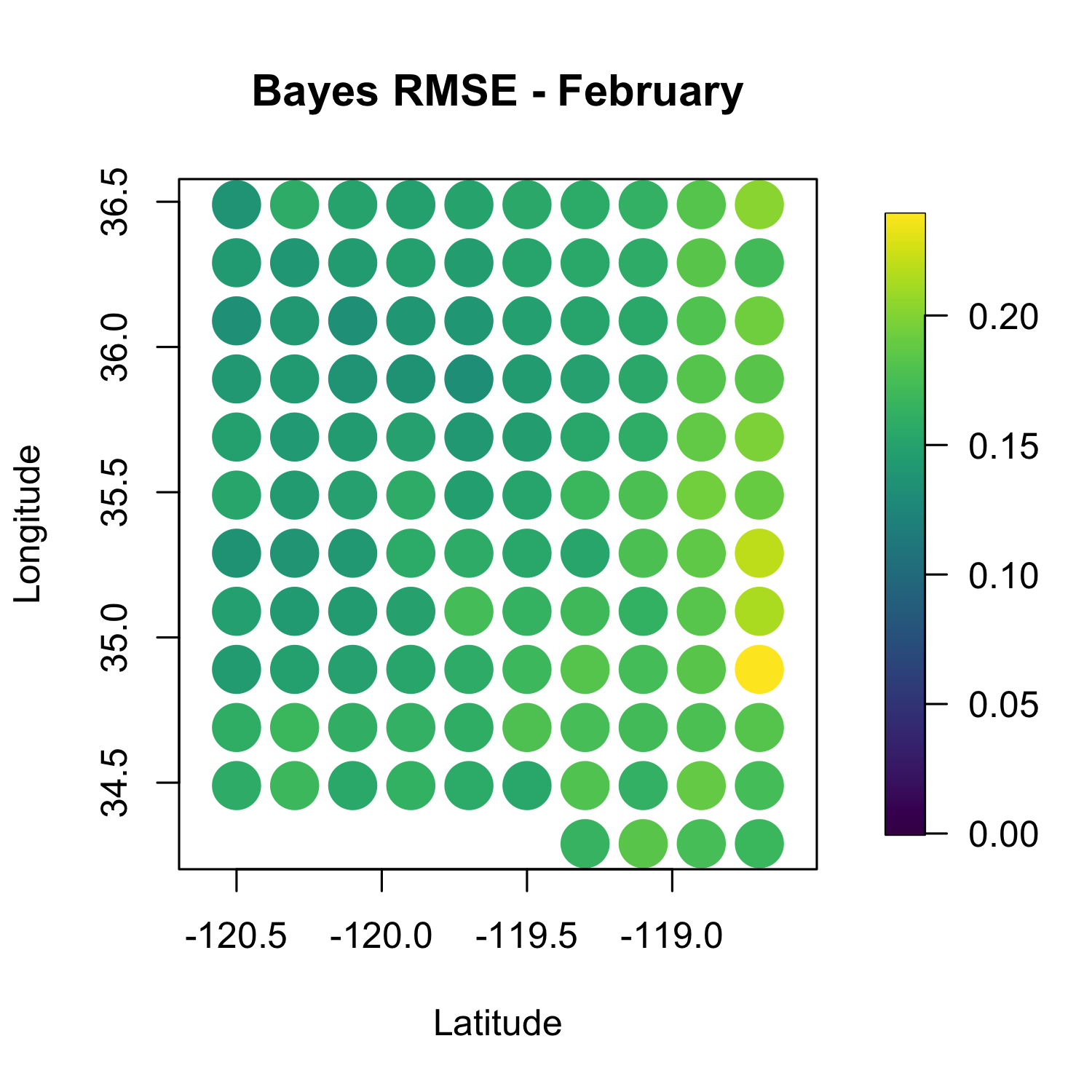}
    \includegraphics[width=.22\textwidth]{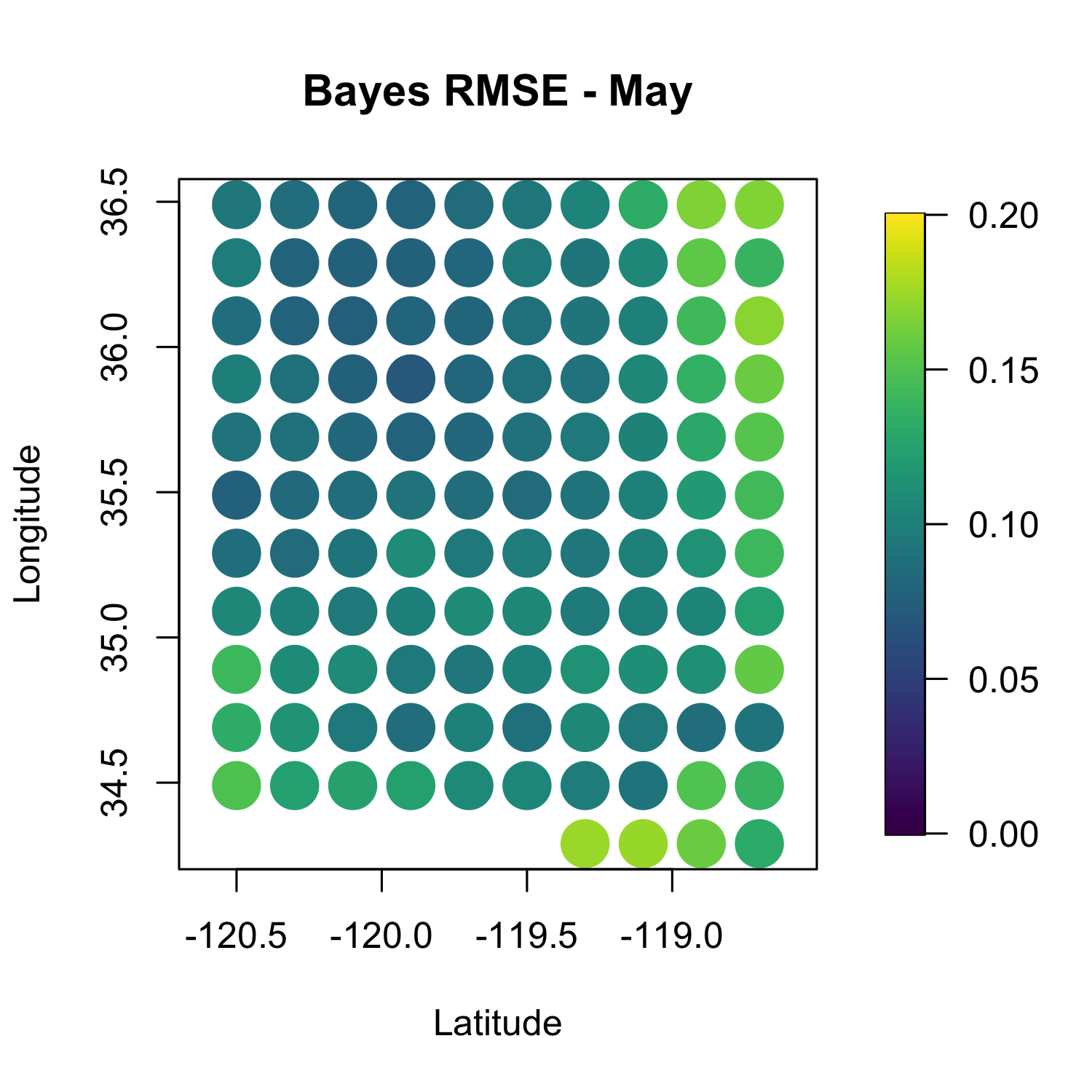}
    \includegraphics[width=.22\textwidth]{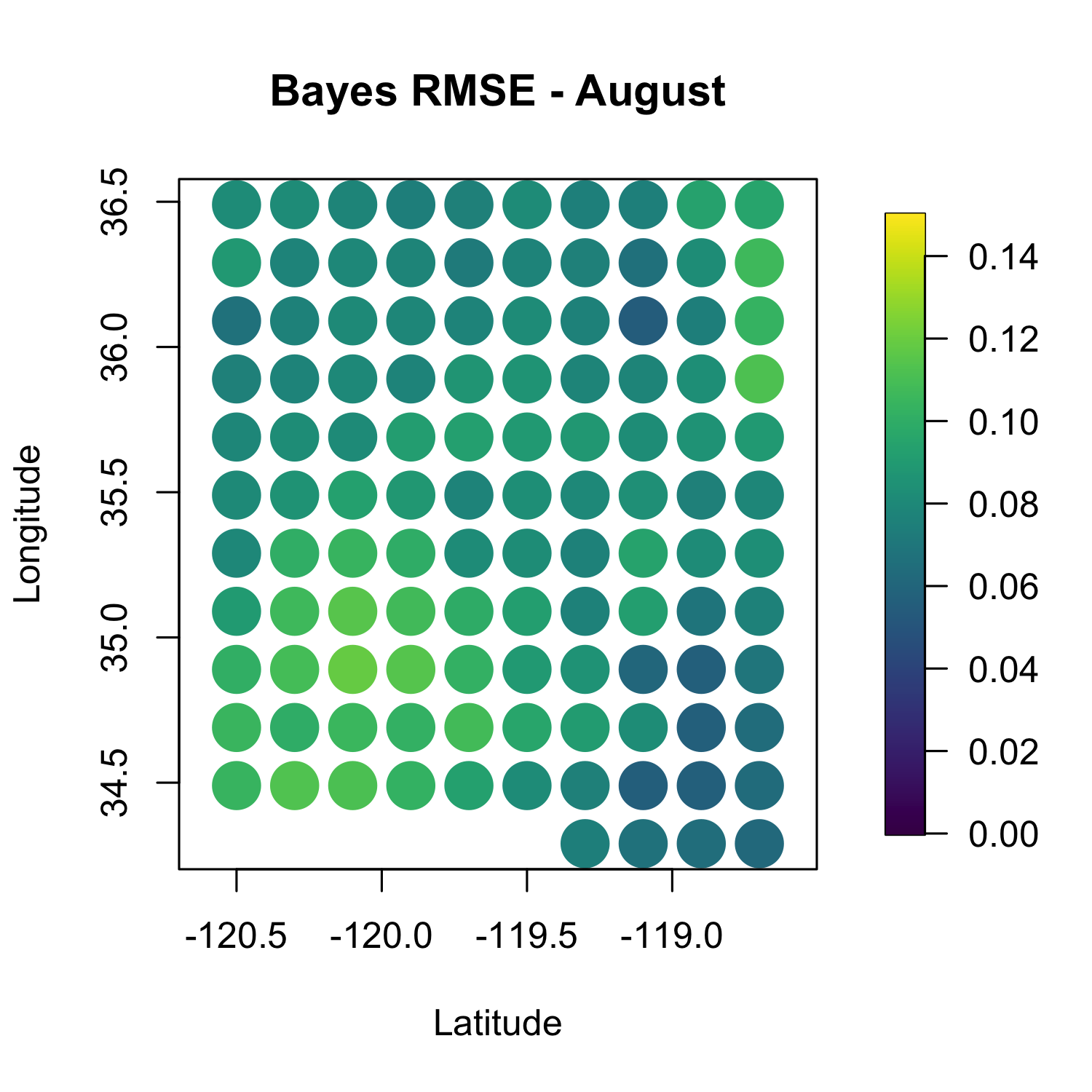}
    \includegraphics[width=.22\textwidth]{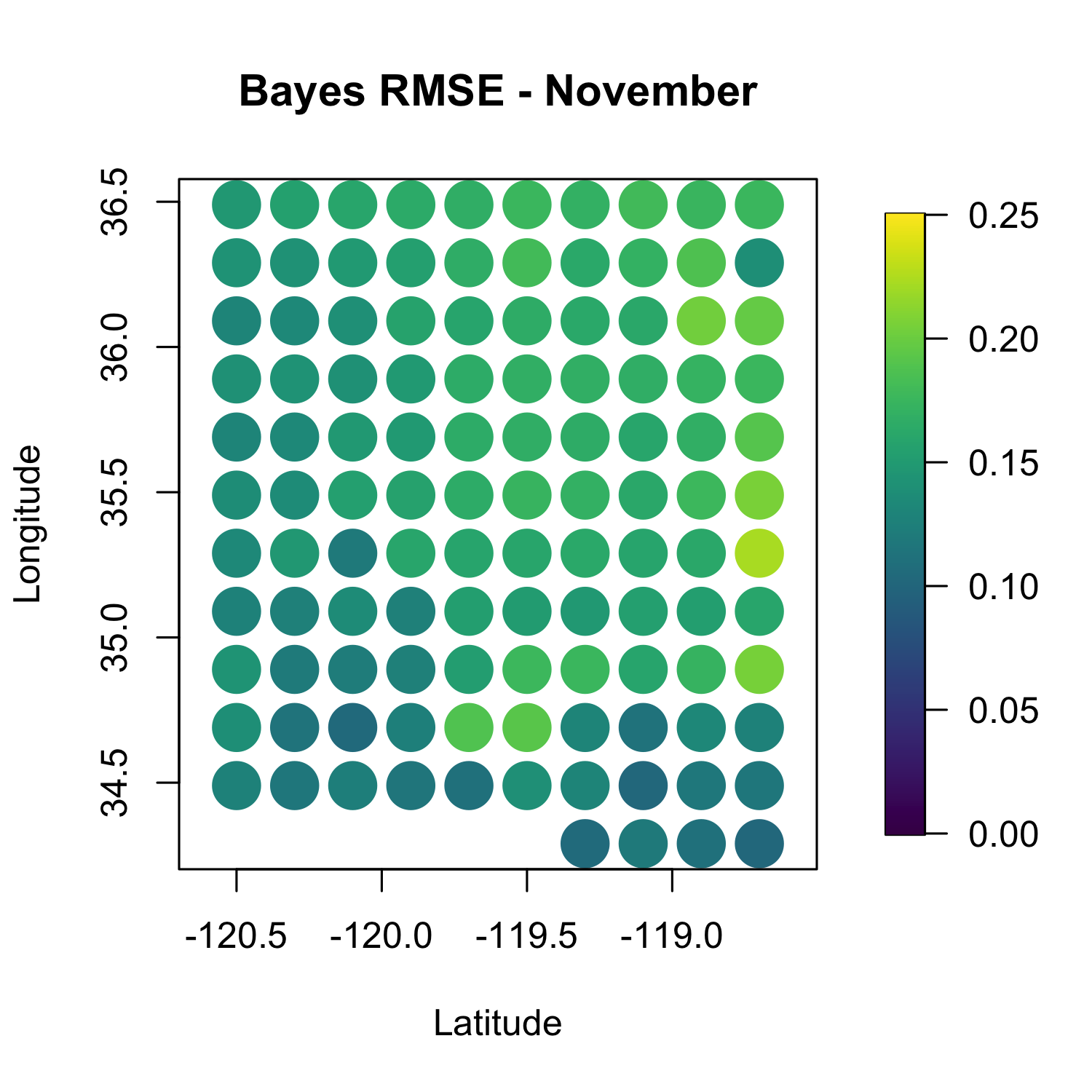}
    \includegraphics[width=.22\textwidth]{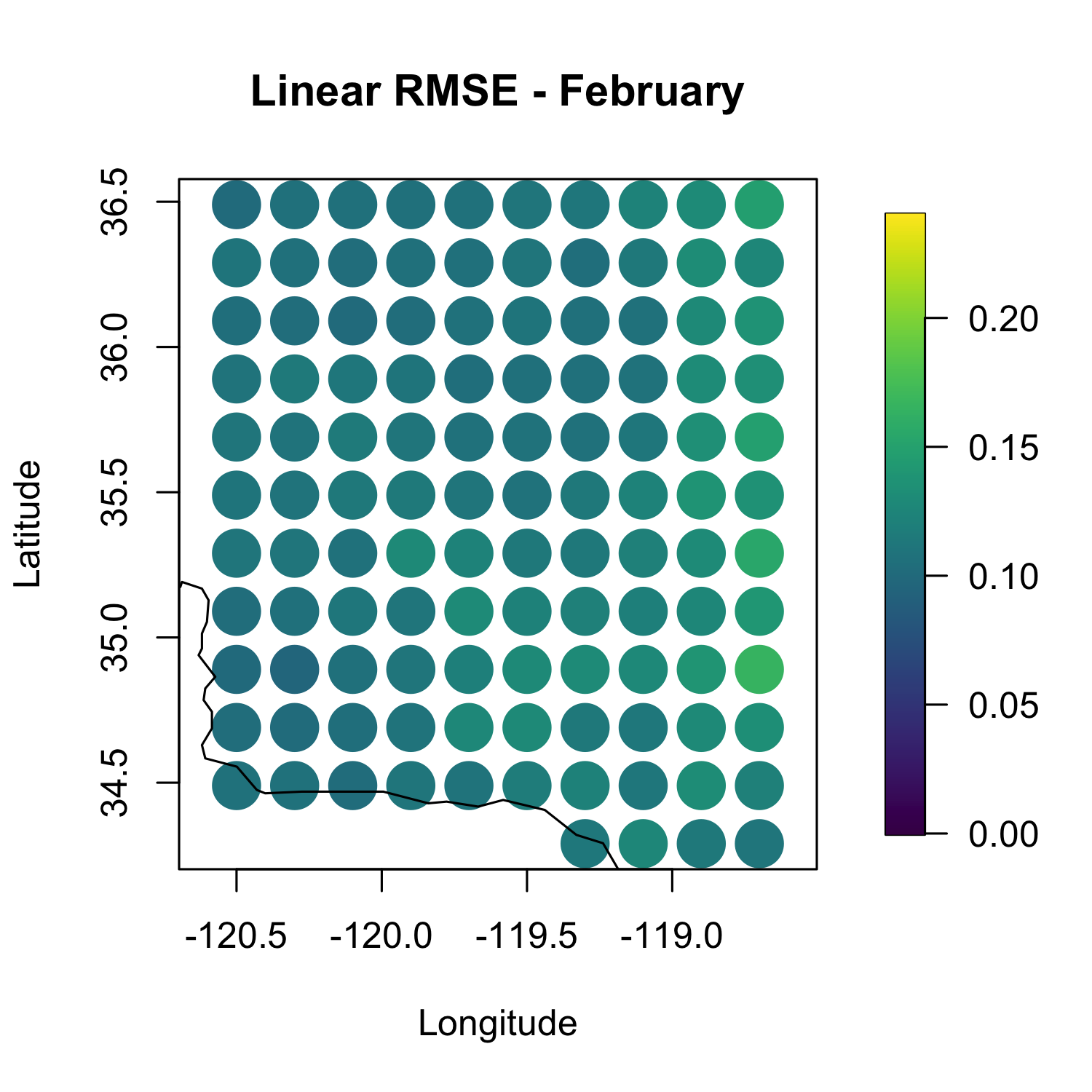} 
    \includegraphics[width=.22\textwidth]{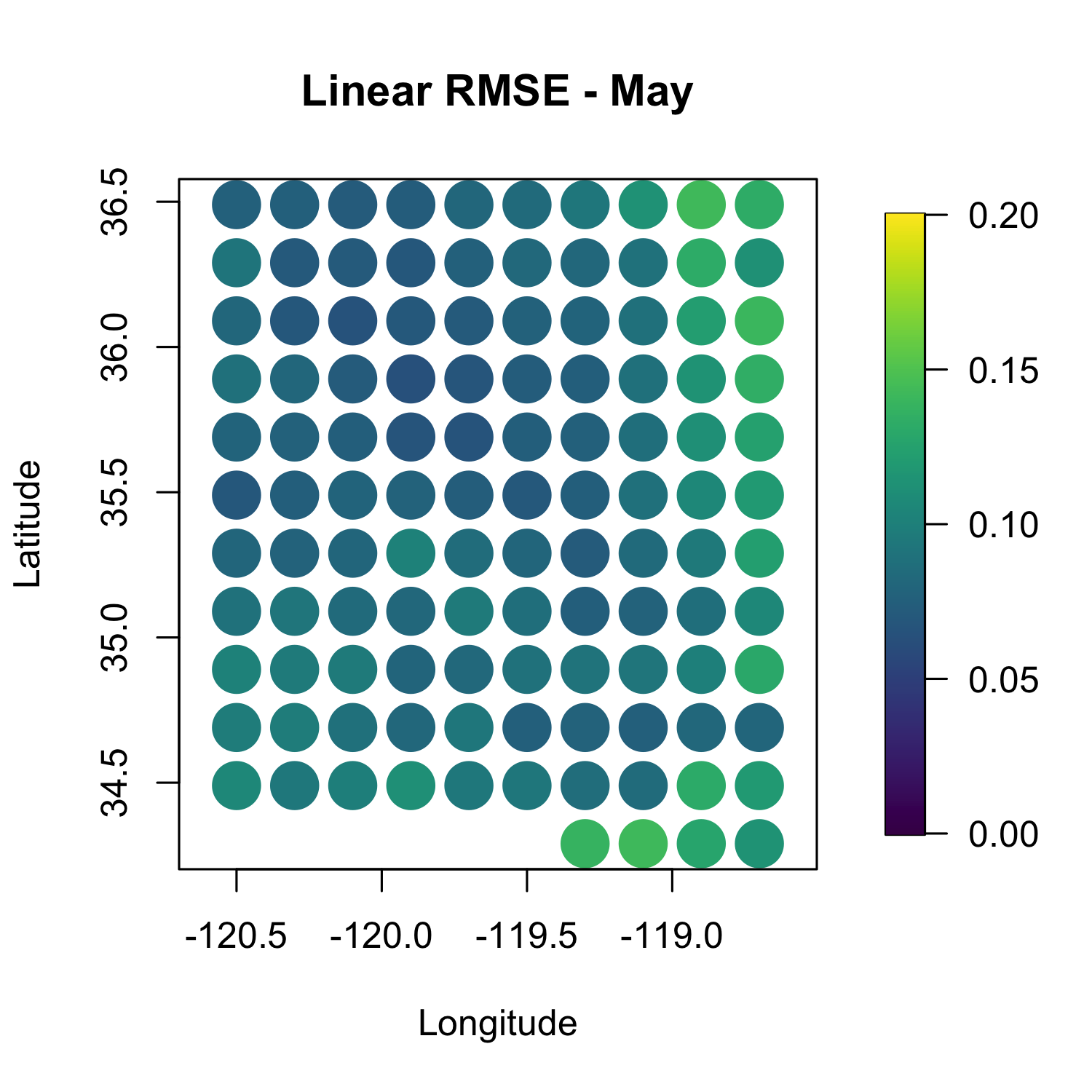} 
    \includegraphics[width=.22\textwidth]{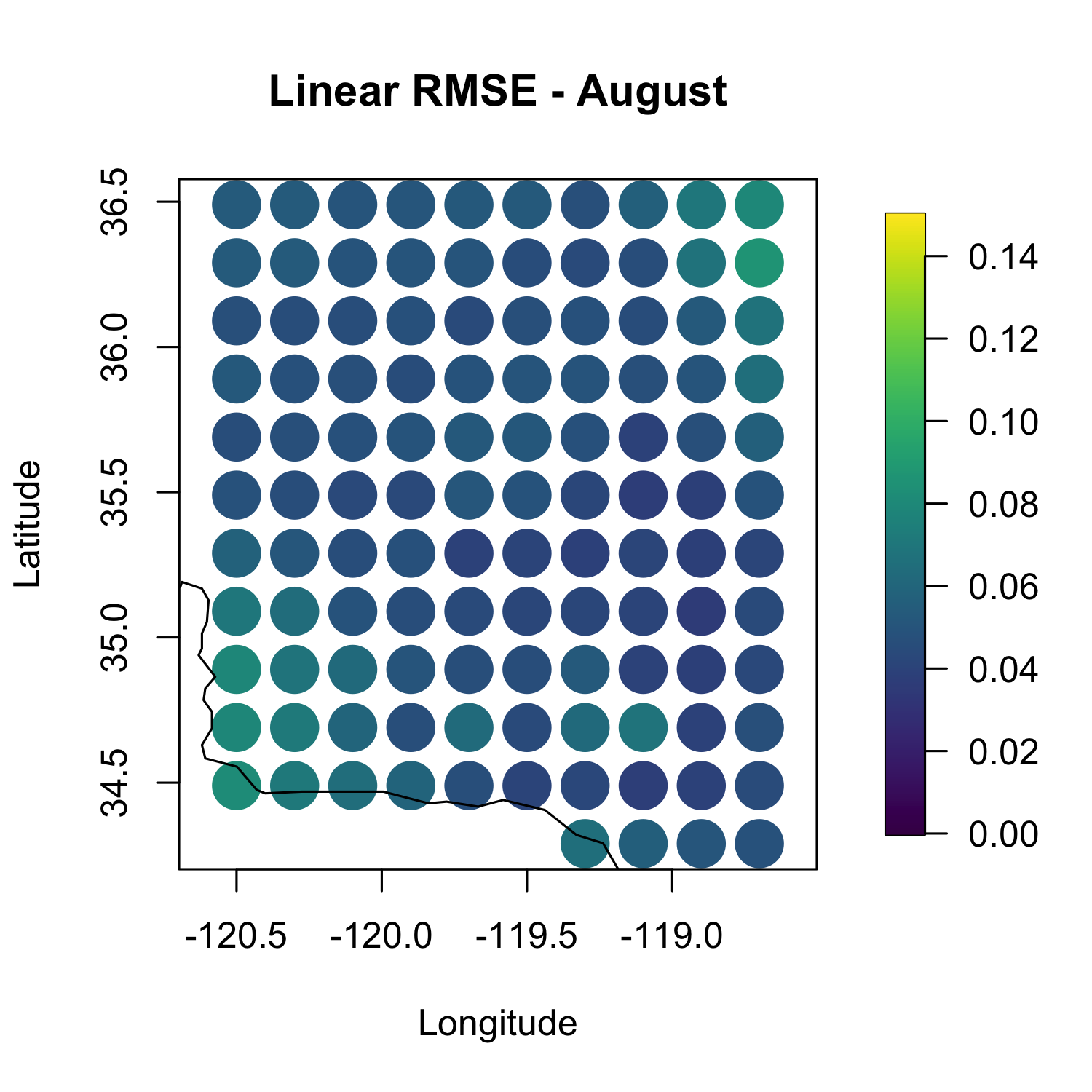} 
    \includegraphics[width=.22\textwidth]{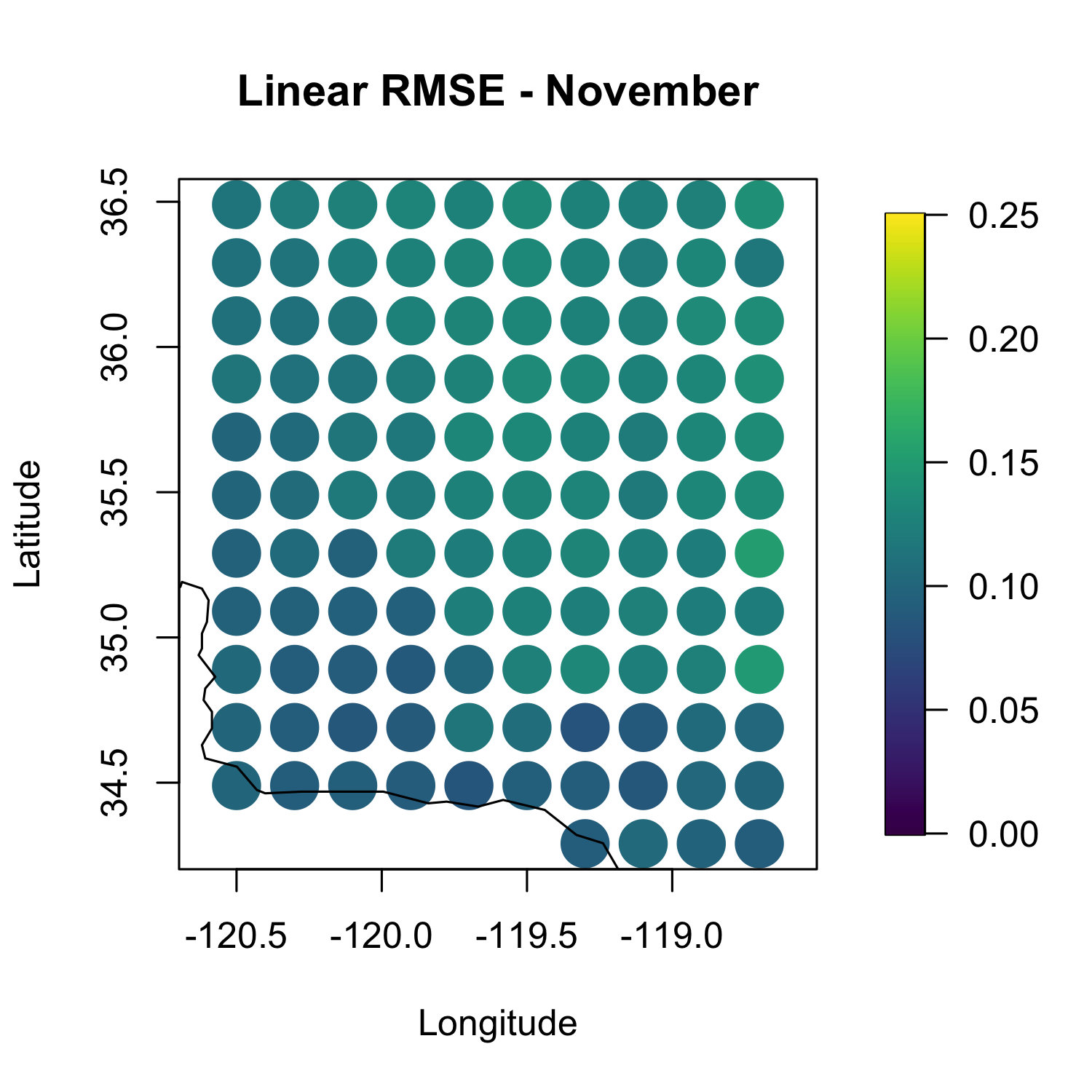}
    \caption{A comparison of the RMSE values for out of sample prediction for the four months considered. Bayesian model results are on the top row and naive regridding results the bottom row. The RMSE is generally higher for the Bayesian results and in November for both models.}
    \label{fig:rmse_comparison}
\end{figure}

Similarly, the RMSE between the predicted GHI and the true GHI is lower across the study domain for August than it is for November in both the naive regridding model and the BHM, indicating better predictions for the summer month over the winter month. This is shown in Figure~\ref{fig:rmse_comparison}. This finding may reflect a characteristic of seasonal solar radiation. Incoming solar radiation during summer months typically has lower standard deviation when considered on a monthly or seasonal basis than in winter in California, indicating that there is less variability in day types (i.e. cloudy versus sunny) or amount of incoming solar radiation during the summer compared to the winter. Therefore, it makes sense that predictions have a lower RMSE in the summer months as the covariables and response have less variability during that season. The RMSE values are also lower for the naive regridding than they are for the BHM across the four months shown. When regridding uncertainty is taken into account, the predicted GHI values have a higher error than when prediction is done directly without considering regridding uncertainty. This is an interesting finding in that it suggests doing prediction directly without considering any uncertainty may produce more accurate point predictions but regridding uncertainty contributes additional variability to the final point estimates as seen in the BHM.

\section{Conclusion}
\label{sec:conclusion}

This study analyzes the uncertainty in regridding of spatial data from climate models, which is often the first step in multi-model climate analysis. Solar radiation data is regridded from its native grid, using kriging with an exponential covariance function and a loglinear transformation,  to the same grid as the NSRDB. Second, we implement a BHM to estimate linear model weights while incorporating the uncertainty associated with the regridding step. Finally, we compare the two and provide an additional simulation study in Appendix~\ref{sec:appendix_a1}. The naive regridding model coefficient estimates were found to be within range of the posterior distributions of the model coefficients in most cases. Seasonally, the month of August produced a mismatch between the naive regridding coefficient and the posterior distribution for the WRF RCM forced by ERA-Interim. In particular, we saw that resulting coefficient estimates in this month for the WRF were higher in the naive method than BHM. This suggests that when regridding uncertainty is taken into account, there is a smaller increase in the WRF data for unit increases in the NSRDB, or that the regridding uncertainty may result in less bias from the WRF in this particular case.

It was found that the posterior coverage for test data for the simulated fields were similar to the naive regridding estimates for the months of August and November. This suggests that when taking into account the regridding uncertainty of the simulated fields and the model parameters themselves, the true value of solar radiation in this case is still likely to be covered by the 95\% credible interval. Therefore if the conditional mean of the regridded field were taken for ground truth, as it often is, downstream effects of regridding on modeling appear to be minimal in the case of solar radiation. However, the BHM had higher RMSE values than the naive regridding models in the months considered indicating that the addition of the regridding uncertainty increased prediction error for out of sample prediction. It is important to note that the naive regridding coefficient estimates give good predictions but are not appropriate to assess the model biases directly since model biases are dependent on regridding.

Finally, this analysis serves as a framework for understanding regridding effects within the context of solar radiation. While this study did not find situations where the BHM regridding consistently outperformed the naive regridding method, we note that this analysis revolves around the chosen variable: GHI. It has been shown that the chosen regridding method has an impact on the extremes of distributions (\cite{mcginnis2010interpolation}), however extremes are not central to solar radiation. A future analysis applying the BHM regridding method to climate variables where extremes of the data are more widely studied, such as precipitation or temperature, may yield different results and provide an example where the method proposed in this paper might show higher uncertainty in downstream modeling. Additionally, this study takes into account a single type of regridding (kriging with an exponential covariance) and this analysis could be extended to other types of interpolation to understand downstream effects of those particular methods. 







\newpage
\appendix

\section{Simulation Study}     
\label{sec:appendix_a1}

\appendixfigures

Here we implement a short simulation study which highlights some of the main differences, as well as similarities,  between the naive method and the BHM.

\subsection{Simulation Study Setup}

For this simulation study, we utilize the same grids from the original regional climate models. The full grids used in the original study are shown in the top row of Figure~\ref{fig:sim_study_grids} with the RCM grids in grey and NSRDB grid over California in blue. The magenta points represent the subsetted area used in the regridding study, which will be also be used in this short simulation study. The bottom row of Figure~\ref{fig:sim_study_grids} shows the grids used for this simulation study. The grey and blue dots show the ``true" grids which include both the RCM and the NSRDB grid. The magenta points are again the simulation study area on the NSRDB grid. This will be explained more fully in the next section. We added a 2 degree buffer around the final study domain in magenta to avoid any edge effects in the simulation and regridding steps.

\subsection{Simulation Study Steps}

\begin{figure}
    \centering
    \includegraphics[width=\textwidth]{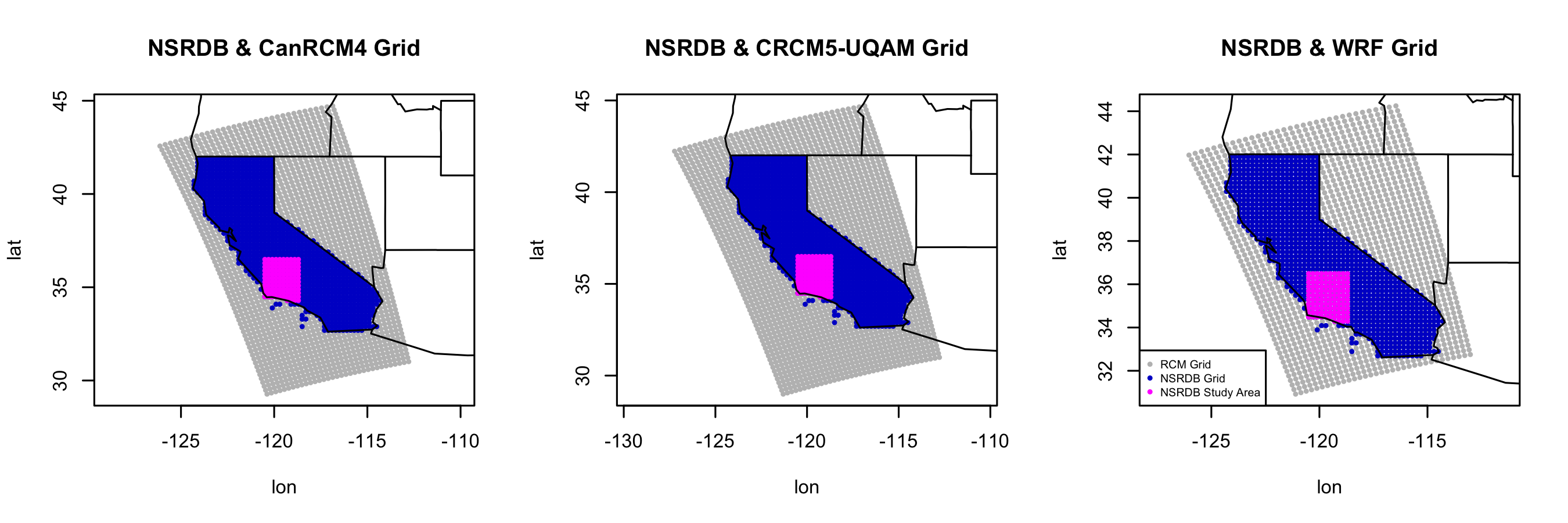}
    \includegraphics[width=\textwidth]{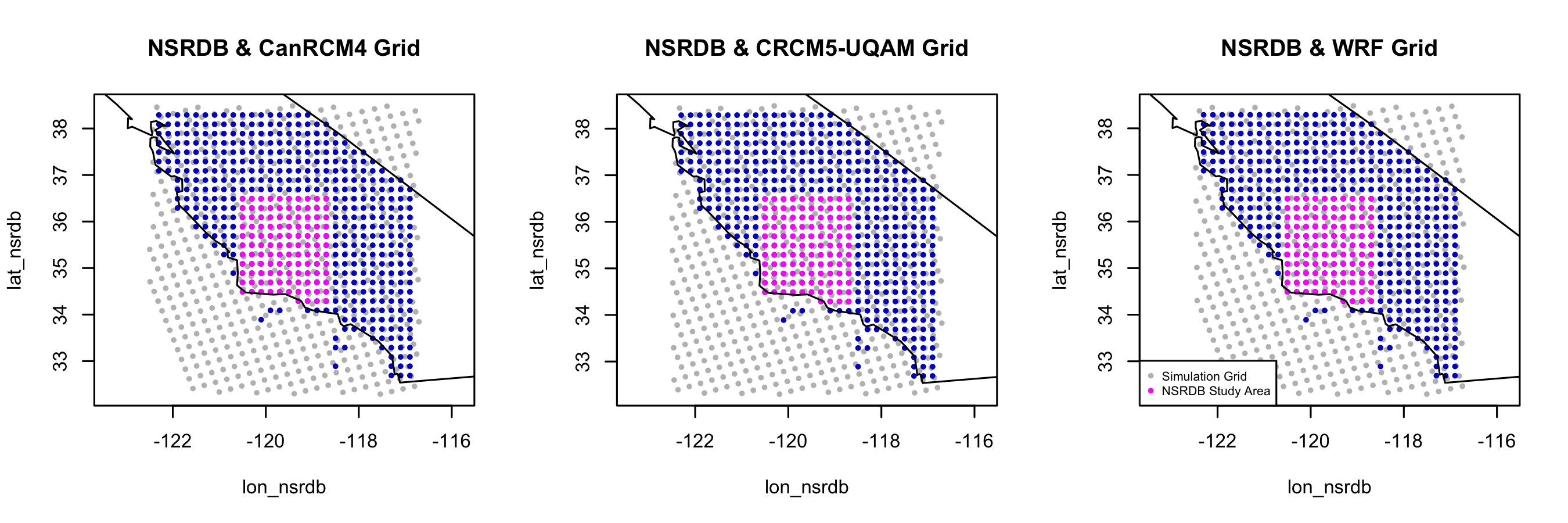}
    \caption{Top: original RCM grid for the three RCMs used in the regridding study plotted with NSRDB grid in blue. Magenta points represent the study area from the regridding study in the main text. Bottom: subsetted grids used in the simulation study.}
    \label{fig:sim_study_grids}
\end{figure}

The simulation study follows the steps:

\begin{enumerate}
    \item Simulate ground truth data 100 times on the combined grids shown in Figure \ref{fig:sim_study_grids}. This is done using the Cholesky decomposition to result in an exact simulation of the Gaussian Process given fixed covariance arguments. That is, for a fixed $\sigma^2$ and $\theta$, the process has the following covariance function
\begin{equation}
  C_\nu(d) = \sigma^2\frac{2^{1-\nu}}{\Gamma(\nu)}\left( \sqrt{2\nu} \frac{d}{\theta}\right)^\nu K_\nu\left( \sqrt{2\nu} \frac{d}{\theta}\right)
\end{equation}

for a distance $d$ between two grid points. The covariance arguments that were estimated in fitting a spatial model to the RCM data for the month of August in the original study were again used here to create a realistic spatial process. This results in 3 sets of combined RCM and NSRDB data that follow the same defined GP. Each of the 100 fields represents a ``day'' of data. Since these days were assumed to be independent in our original study, we maintain this independence in the simulation here. The parameters used for each Gaussian process are in Table~\ref{table:sim_gp_params}.

    \item Subset the grey points in the bottom row of Figure~\ref{fig:sim_study_grids} to only the points that fall on the original RCM grid, in grey in the top row of Figure~\ref{fig:sim_study_grids}. The ``true'' NSRDB observations are generated by weighting each RCM grid in a linear model by location:
\begin{equation}
  \boldsymbol{y}^i = \boldsymbol{\beta}_0 + \sum_{j = 1}^3 \beta_jX_j^i + \varepsilon^i
\end{equation}
where $\varepsilon^i \sim N(0, \gamma^i)$ and $\gamma^i$ is fixed as the estimated variance of the residuals from the regridding study in primary manuscript for the month of August. The estimated coefficients from the naive method for the same month from the original study are used here as weights for each RCM to create a realistic ``observation'' set on the magenta points, or the NSRDB study area.
    \item  Regridding and model fitting is done for each method in the following way:
    \begin{enumerate}
        \item \textit{Bayesian Regridding}: Conditionally simulate each day, or field, 100 times onto the NSRDB Study Area from the RCM grids. For each simulation, run the BHM as described in Section~\ref{sec:approx_bhm} and take 50 draws from the posterior distribution for each coefficient resulting in a posterior distribution for each location and for each coefficient.
        \item \textit{Naive Regridding}: Regrid once from the RCM grid to the NSRDB grid using Kriging. Then estimate a linear model for each location across the 100 days, or fields.
    \end{enumerate}
    \item Compare the coefficient estimates and standard errors of the coefficients from the naive regridding and the BHM regridding method. 
\end{enumerate}

\subsection{True Coefficients}

\begin{figure}
    \centering
    \includegraphics[width=\textwidth]{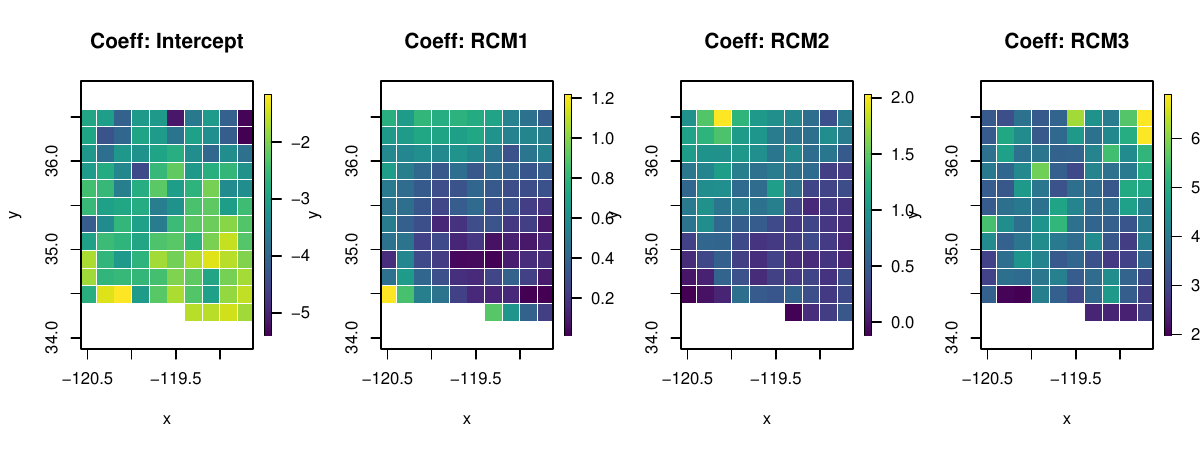}
    \caption{Value of $\beta$ fixed at each location in the study area.}
    \label{fig:sim_study_true_coeffs}
\end{figure}

In step 2, the true response, representing the observations from the NSRDB, are generated by weighting each RCM at each location in the study area across all 100 days. These coefficients are set by taking the average $\hat{\beta}$ from the BHM in the original manuscript study. The true values for the intercept and each RCM at each location are shown in Figure~\ref{fig:sim_study_true_coeffs}. The weights correspond to the weights estimated for the CanRCM4, CRCM5-UQAM and WRF. The WRF had the highest weight between the three RCMs in the original study.

\subsection{Results}

\begin{figure}
    \centering
    \includegraphics[width=\textwidth]{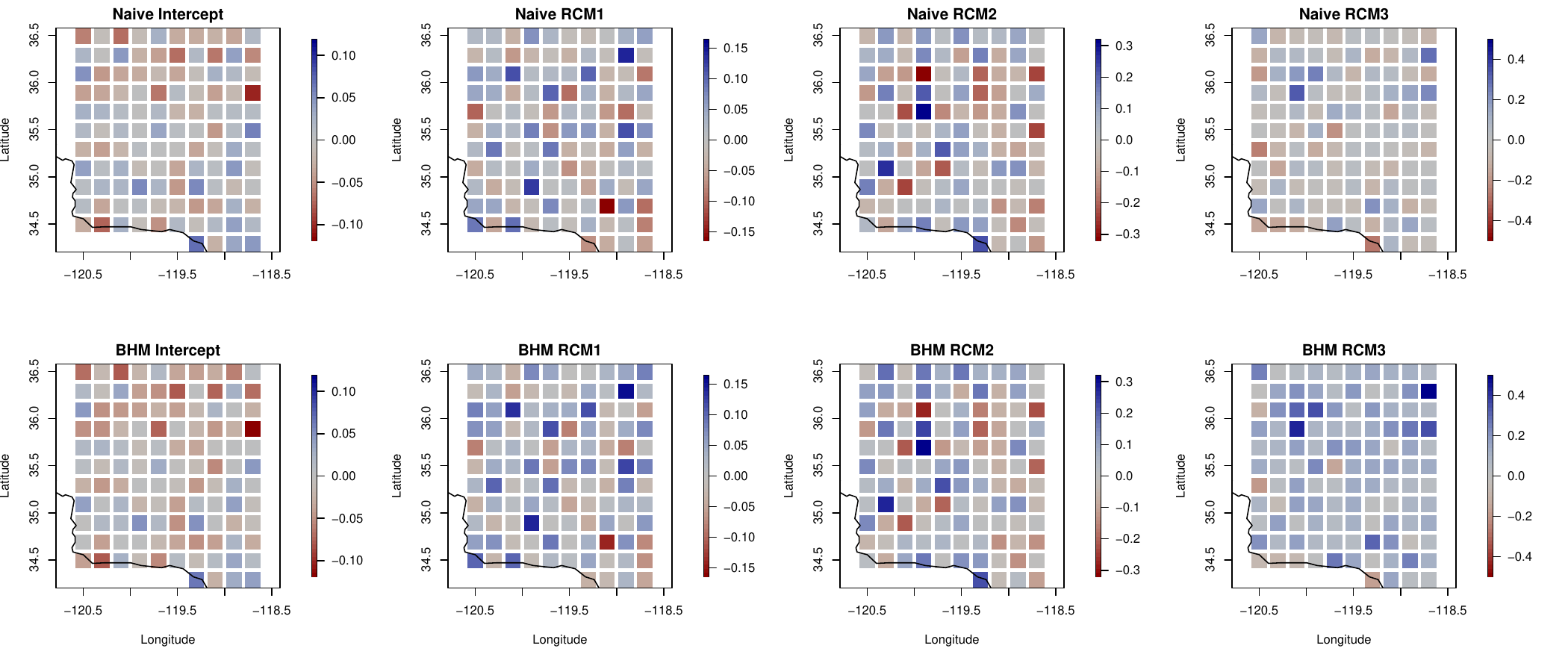}
    \vspace{1em}
    \includegraphics[width=\textwidth]{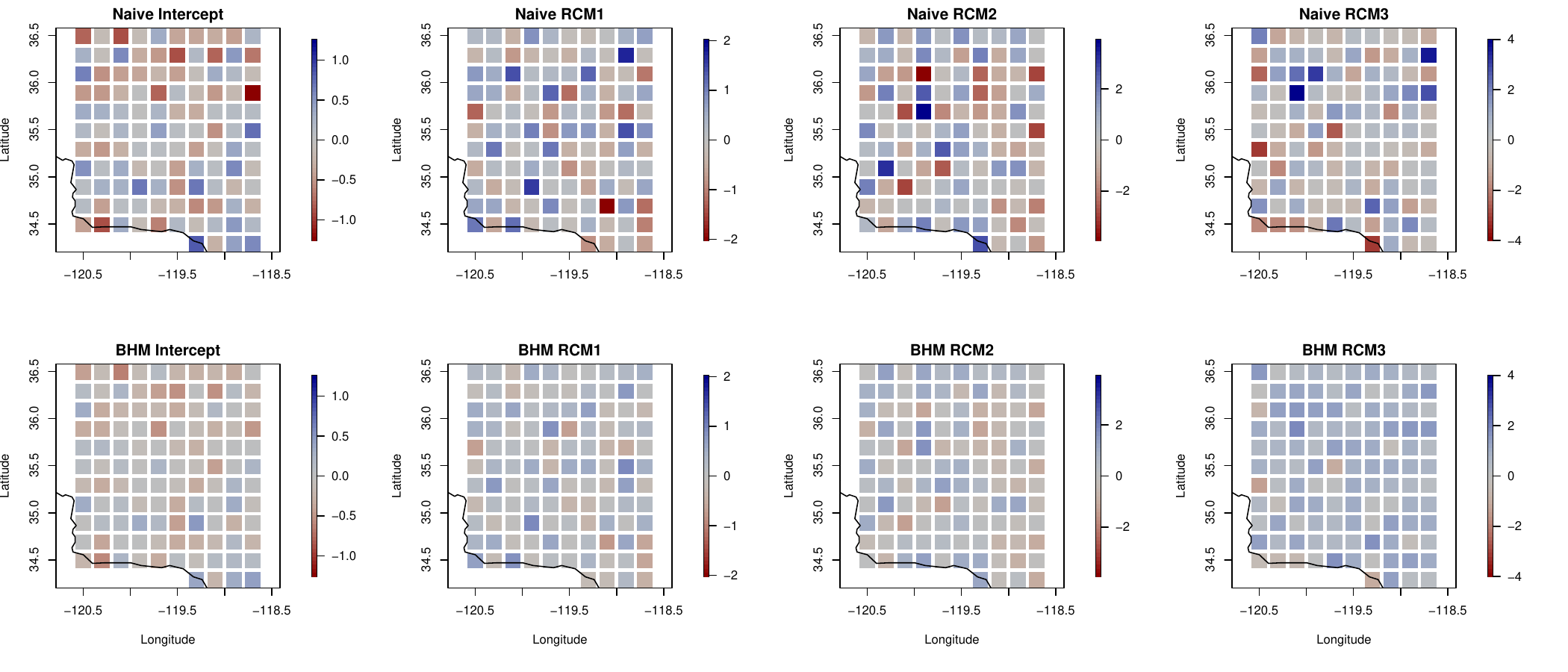}
    \caption{Top: bias between the resulting coefficient estimates from the two methods and the true coefficients. Bottom: bias between the resulting coefficient estimates normalized by the RMSE between the true and estimated coefficients from each method.}
    \label{fig:sim_study_coeff_bias}
\end{figure}

The raw difference relative to the RMSE between the estimated $\hat{\beta}$ from the linear model and mean of the posterior distribution for the BHM are shown in Figure~\ref{fig:sim_study_coeff_bias}. Generally, relative to the overall variation of the error, the difference between the true and estimate coefficients is smaller for the BHM and higher for the naive method. This indicates that the regridding uncertainty decreases the raw error in the estimated weights. We also see that for the largest weight (RCM3) the BHM slightly \textit{underestimates} $\beta_3$ but the naive method sees larger errors. In general we can say that the bias between the estimates from the BHM is lower than that of the naive regridding method.

We can also look that raw error without normalizing by the RMSE, seen in Figure \ref{fig:sim_study_coeff_bias}. This figure shows similar errors between the two methods compared to the true coefficients. The exception is seen in the coefficient estimate for RCM3, which is slightly overestimated by the BHM when considering the raw difference compared to the RCM, which slightly underestimates or is closer to the true coefficient.

\begin{figure}
    \centering
    \includegraphics[width=\textwidth]{ 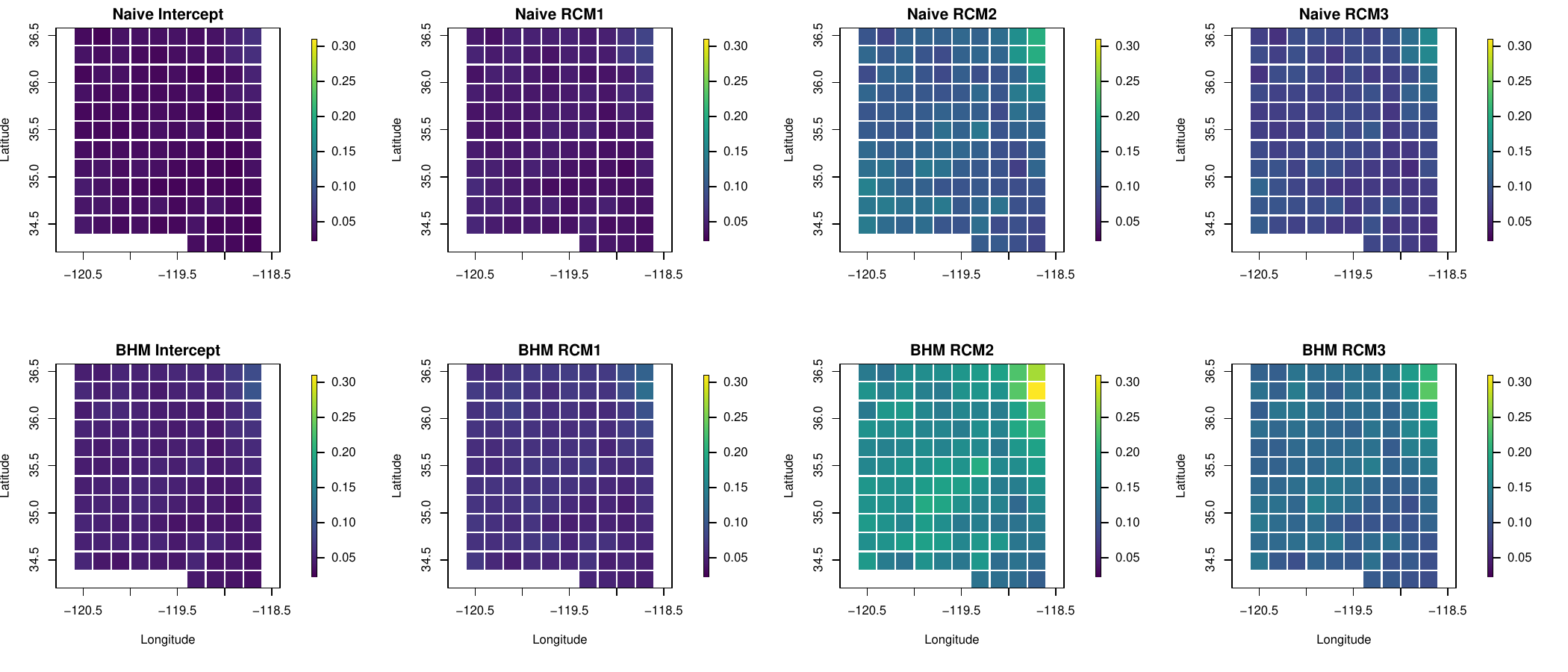}
    \caption{Standard error of the estimated coefficients at each location in the simulation study area.}
    \label{fig:coeff_std_error}
\end{figure}

One interesting finding is that the coefficient standard error is larger for the posterior distribution of the BHM estimates than it is for the naive coefficient estimates. This is seen in Figure \ref{fig:coeff_std_error}, where the top row shows the standard error for the naive regridding coefficient estimates and the bottom row shows the posterior standard deviations of the BHM estimates. The coefficient standard error for the BHM estimates is on average 46\% larger than it is for the naive method. This signifies that incorporating the regridding uncertainty results in a slightly wider range of potential weights for the RCMs in predicting the observed values. However, the difference in the standard error is not large and because the posterior mean of the BHM coefficient estimates and the naive regridding estimates are very similar, this result may not have much of an effect in this application. 

\subsection{Simulation Study Conclusions}

This simulation study highlights some important differences between the naive regridding and BHM regridding methods. In particular, the biases between the true coefficients (weights) and the estimated weights is smaller relative to the error variance for the BHM and larger for the naive method. This indicates that when regridding uncertainty is taken into account, the BHM is more precise in estimating the RCM weights. This finding is further emphasized in the standard error of the coefficients for the two methods. This value is lower for the BHM than it is in the naive method, though the difference in standard error between the two methods is small. Overall we can conclude that taking regridding uncertainty into account results in more precise estimates for model weights when predicting observed values based on climate model output. This precision provides increased understanding in which climate model may have higher importance in predicting observed outcomes and, therefore, future projections for solar radiation.

\appendixtables
\begin{table}
\begin{center}
\begin{tabular}{|l|c|c|}
\hline
     & Marginal Variance & Range \\
     \hline
     \hline
      RCM1 (CanRCM4) & 0.58 & 8.02 \\
      RCM2 (CRCM5-UQAM) & 0.095	& 2.81\\
      RCM3 (WRF) & 0.16	& 4.62 \\
      \hline
\end{tabular}
\vspace{1em}
\caption{Covariance arguments used to simulated RCM data. All covariance functions are a Mat\'ern with smoothness = 1. No measurement error is included.}
\label{table:sim_gp_params}
\end{center}
\end{table}


\clearpage

\section{Regridding Coefficient Estimates}
\label{sec:appendix_b1}

 \begin{figure}[H]
 \centering
   \rotatebox{90}{\includegraphics[height = 0.7\textheight]{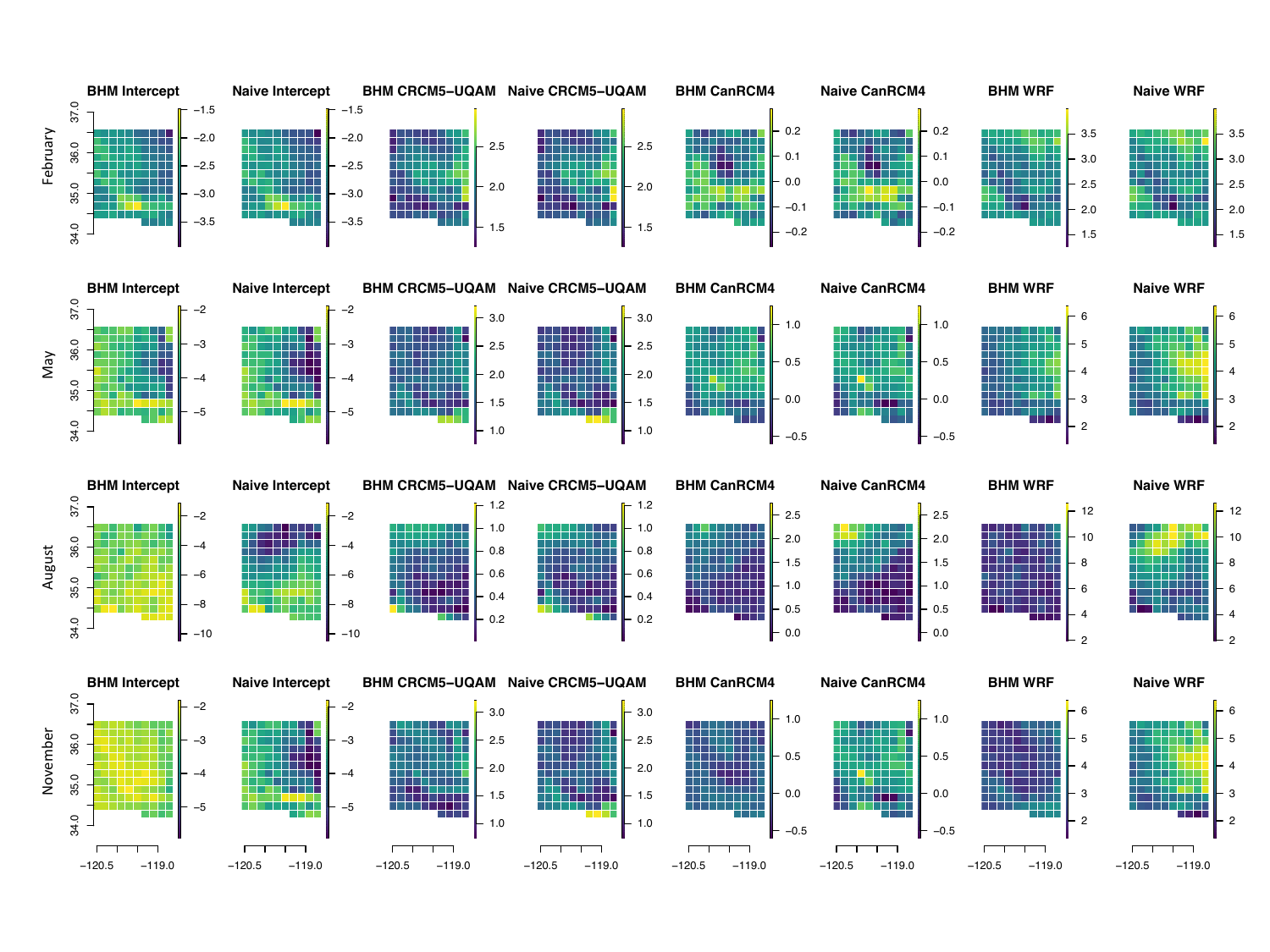}}
   \caption{Coefficient estimates for the intercept and each RCM for each month considered between the two methods. }
  \label{fig:beta_hat_all_seasons}
 \end{figure}

 \begin{figure}
 \centering
   \rotatebox{90}{\includegraphics[height = 0.7\textheight]{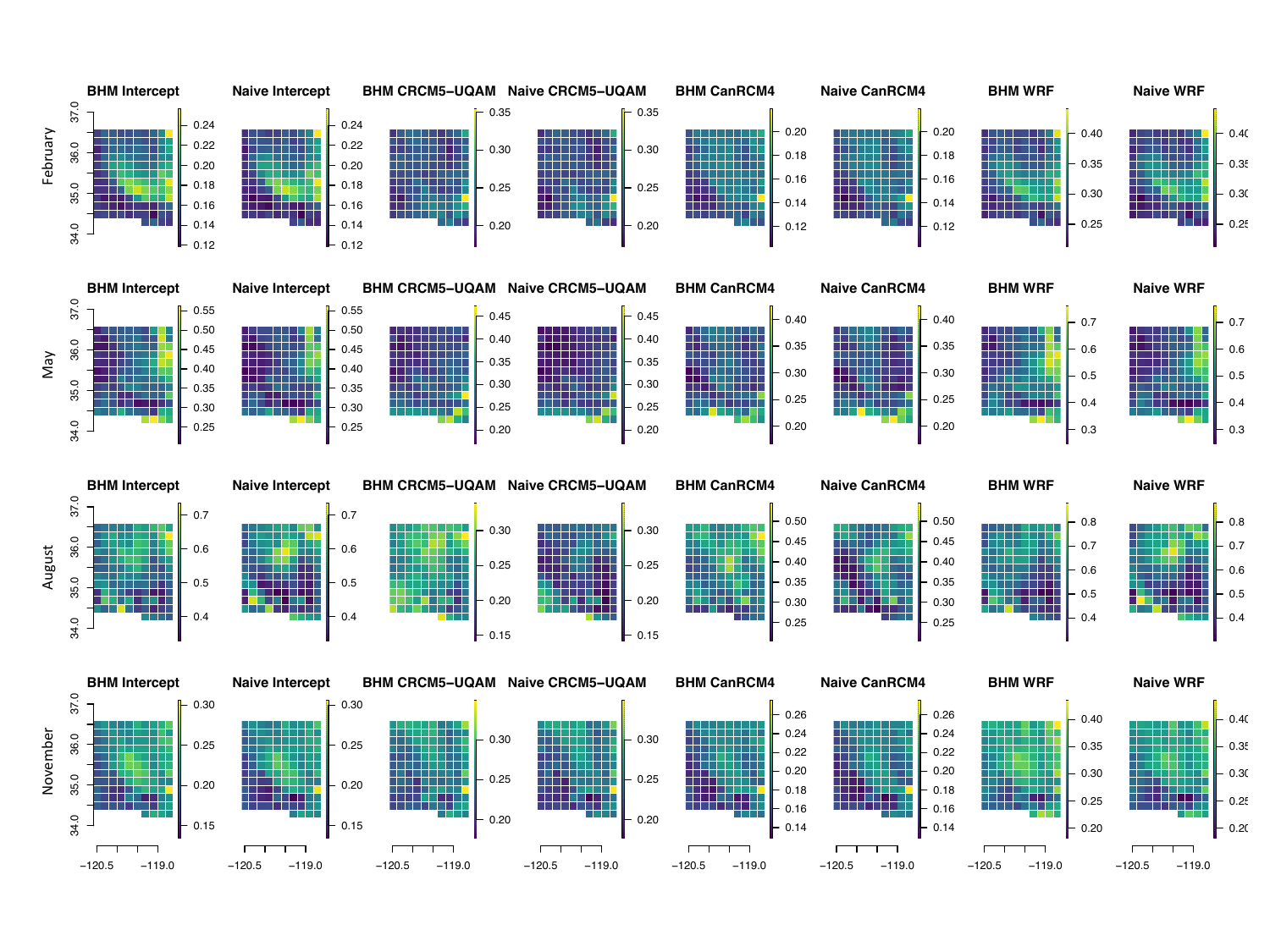}}
  \caption{Standard error of the coefficient estimates for the intercept and each RCM for each month considered between the two methods.}
  \label{fig:sd_beta_all_seasons}
 \end{figure}

\noappendix       







\clearpage

\authorcontribution{MDB conducted all statistical analyses and wrote the manuscript; SB and DN contributed to the manuscript and advised on all statistical analyses; MS, AH, and YX provided expertise in the application to solar radiation.} 

\competinginterests{The contact author has declared that none of the authors has any competing interests.} 

\financialsupport{Funding provided by the U.S. Department of Energy Office of Energy Efficiency and Renewable Energy Solar Energy Technologies Office.}


\begin{acknowledgements}
This work was authored by the National Renewable Energy Laboratory, operated by Alliance for Sustainable Energy, LLC, for the U.S. Department of Energy (DOE) under Contract No. DE-AC36-08GO28308. The views expressed in the article do not necessarily represent the views of the DOE or the U.S. Government. The U.S. Government retains and the publisher, by accepting the article for publication, acknowledges that the U.S. Government retains a nonexclusive, paid-up, irrevocable, worldwide license to publish or reproduce the published form of this work, or allow others to do so, for U.S. Government purposes.
\end{acknowledgements}








\clearpage
\bibliographystyle{copernicus}
\bibliography{example.bib}

\begin{thebibliography}{19}
\providecommand{\natexlab}[1]{#1}
\providecommand{\url}[1]{{\tt #1}}
\providecommand{\urlprefix}{URL }
\expandafter\ifx\csname urlstyle\endcsname\relax
  \providecommand{\doi}[1]{https://doi.org/\discretionary{}{}{}#1}\else
  \providecommand{\doi}{https://doi.org/\discretionary{}{}{}\begingroup
  \urlstyle{rm}\Url}\fi

\bibitem[{Accadia et~al.(2003)Accadia, Mariani, Casaioli, Lavagnini, and
  Speranza}]{accadia2003sensitivity}
Accadia, C., Mariani, S., Casaioli, M., Lavagnini, A., and Speranza, A.:
  Sensitivity of precipitation forecast skill scores to bilinear interpolation
  and a simple nearest-neighbor average method on high-resolution verification
  grids, Weather and forecasting, 18, 918--932, 2003.

\bibitem[{Berndt and Haberlandt(2018)}]{berndt2018spatial}
Berndt, C. and Haberlandt, U.: Spatial interpolation of climate variables in
  Northern Germany—Influence of temporal resolution and network density,
  Journal of Hydrology: Regional Studies, 15, 184--202, 2018.

\bibitem[{Chandler et~al.(2022)Chandler, Barnes, Brierley, and
  Alegre}]{chandler2022regridding}
Chandler, R., Barnes, C., Brierley, C., and Alegre, R.: Regridding and
  interpolation of climate data for impacts modelling--some cautionary notes,
  Tech. rep., Copernicus Meetings, 2022.

\bibitem[{Cressie and Wikle(2011)}]{Cressie2011StatisticsFS}
Cressie, N. and Wikle, C.~K.: Statistics for Spatio-Temporal Data, 2011.

\bibitem[{Cressie(1996)}]{cressie1996change}
Cressie, N.~A.: Change of support and the modifiable areal unit problem, 1996.

\bibitem[{Diaconescu et~al.(2015)Diaconescu, Gachon, and
  Laprise}]{diaconescu2015remapping}
Diaconescu, E.~P., Gachon, P., and Laprise, R.: On the remapping procedure of
  daily precipitation statistics and indices used in regional climate model
  evaluation, Journal of Hydrometeorology, 16, 2301--2310, 2015.

\bibitem[{Ensor and Robeson(2008)}]{ensor2008statistical}
Ensor, L.~A. and Robeson, S.~M.: Statistical characteristics of daily
  precipitation: comparisons of gridded and point datasets, Journal of Applied
  Meteorology and Climatology, 47, 2468--2476, 2008.

\bibitem[{Finley et~al.(2013)Finley, Banerjee, and Gelfand}]{finley2013spbayes}
Finley, A.~O., Banerjee, S., and Gelfand, A.~E.: spBayes for large univariate
  and multivariate point-referenced spatio-temporal data models, arXiv preprint
  arXiv:1310.8192, 2013.

\bibitem[{Gelfand et~al.(2001)Gelfand, Zhu, and Carlin}]{gelfand2001change}
Gelfand, A.~E., Zhu, L., and Carlin, B.~P.: On the change of support problem
  for spatio-temporal data, Biostatistics, 2, 31--45, 2001.

\bibitem[{Handcock and Stein(1993)}]{Handcock1993ABA}
Handcock, M.~S. and Stein, M.~L.: A Bayesian analysis of kriging,
  Technometrics, 35, 403--410, 1993.

\bibitem[{Harris et~al.(2022)Harris, Li, and Sriver}]{harris2022multi}
Harris, T., Li, B., and Sriver, R.: Multi-model Ensemble Analysis with Neural
  Network Gaussian Processes, arXiv [preprint] arXiv:2202.04152, 2022.

\bibitem[{Loghmari et~al.(2018)Loghmari, Timoumi, and
  Messadi}]{loghmari2018performance}
Loghmari, I., Timoumi, Y., and Messadi, A.: Performance comparison of two
  global solar radiation models for spatial interpolation purposes, Renewable
  and Sustainable Energy Reviews, 82, 837--844, 2018.

\bibitem[{McGinnis and Mearns(2021)}]{mcginnis2021building}
McGinnis, S. and Mearns, L.: Building a climate service for North America based
  on the NA-CORDEX data archive, Climate Services, 22, 100\,233, 2021.

\bibitem[{McGinnis et~al.(2010)McGinnis, Mearns, and
  McDaniel}]{mcginnis2010interpolation}
McGinnis, S., Mearns, L., and McDaniel, L.: Effects of Spatial Interpolation
  Algorithm Choice on Regional Climate Model Data Analysis, "Fall Meeting,
  American Geophysical Union, San Francisco", 2010.

\bibitem[{Phillips and Marks(1996)}]{phillips1996spatial}
Phillips, D.~L. and Marks, D.~G.: Spatial uncertainty analysis: propagation of
  interpolation errors in spatially distributed models, Ecological Modelling,
  91, 213--229, 1996.

\bibitem[{Rajulapati et~al.(2021)Rajulapati, Papalexiou, Clark, and
  Pomeroy}]{rajulapati2021perils}
Rajulapati, C.~R., Papalexiou, S.~M., Clark, M.~P., and Pomeroy, J.~W.: The
  perils of regridding: examples using a global precipitation dataset, Journal
  of Applied Meteorology and Climatology, 60, 1561--1573, 2021.

\bibitem[{Rauscher et~al.(2010)Rauscher, Coppola, Piani, and
  Giorgi}]{rauscher2010resolution}
Rauscher, S.~A., Coppola, E., Piani, C., and Giorgi, F.: Resolution effects on
  regional climate model simulations of seasonal precipitation over Europe,
  Climate dynamics, 35, 685--711, 2010.

\bibitem[{Sengupta et~al.(2018)Sengupta, Xie, Lopez, Habte, Maclaurin, and
  Shelby}]{sengupta2018national}
Sengupta, M., Xie, Y., Lopez, A., Habte, A., Maclaurin, G., and Shelby, J.: The
  national solar radiation data base (NSRDB), Renewable and sustainable energy
  reviews, 89, 51--60, 2018.

\bibitem[{Whittemore(1989)}]{whittemore1989errors}
Whittemore, A.~S.: Errors-in-variables regression using Stein estimates, The
  American Statistician, 43, 226--228, 1989.

\end{thebibliography}

\end{document}